\documentclass[aps,pre,reprint,amsmath,amssymb,superscriptaddress]{revtex4-2}

\usepackage{graphicx}
\usepackage{bm}

\DeclareMathOperator{\sech}{sech}

\newcommand{\anorm}{\alpha_{\mathrm{norm}}}
\newcommand{\anormtilde}{\tilde{\alpha}_{\mathrm{norm}}}
\newcommand{\aefflab}{\alpha_{\mathrm{eff}}^{\mathrm{lab}}}
\newcommand{\aefflabtilde}{\tilde{\alpha}_{\mathrm{eff}}^{\mathrm{lab}}}

\usepackage{booktabs}
\usepackage[utf8]{inputenc}
\usepackage{soul}

\usepackage[T1]{fontenc}
\usepackage{xcolor}
\usepackage{pdfcolmk}       
\soulregister\cite7
\soulregister\ref7
\soulregister\eqref7
\soulregister\pageref7

\usepackage{pgfplots}
\pgfplotsset{compat=1.18}

\usepackage[para,flushleft]{threeparttable}

\usepackage{hyperref}

\begin{document}

\title{Admissibility of Solitary Wave Modes in Long-Runout Debris Flows}

\author{Louis-S. Bouchard}
\affiliation{UCLA Department of Chemistry and Biochemistry, 607 Charles E Young Drive East, Los Angeles, CA 90095}

\author{Seulgi Moon}
\affiliation{UCLA Department of Earth, Planetary, and Space Sciences, 595 Charles E Young Drive East, Los Angeles, CA 90095}

\begin{abstract}
Debris flows often exhibit coherent wave structures—shock-like roll waves on steeper slopes and weaker, more sinusoidal dispersive pulses on gentler slopes. Coarse-rich heads raise basal resistance, whereas fines-rich tails lower it; in gentle reaches, small-amplitude pulses can locally transport momentum across low-resistance segments. We focus on this gentle-slope, long-wave, low-amplitude regime, where the base-flow Froude number is order unity.
In this limit, we obtain a Korteweg--de Vries (KdV) reduction from depth-averaged balances with frictional (Coulomb) and viscous–plastic basal options, using a curvature-type internal normal-stress closure in the long-wave small-$k$ regime. Multiple-scale analysis yields effective nonlinear and dispersive coefficients. We also introduce a practical nonlinearity diagnostic that can be computed from observed crest speeds and flow thicknesses. When laboratory-frame crest celerity is available, we estimate an effective quadratic coefficient from the KdV speed–amplitude relation and report its ratio to the shallow-water reference. When only a depth-averaged first-surge speed and thickness are available, we use the same construction to form a velocity-based proxy and note its bias near zero.
A Froude–slope diagram organizes published cases into a steep-slope roll-wave domain and a gentle-slope corridor where KdV pulses are admissible. Numerical solutions of the full depth-averaged model produce cnoidal and solitary waves that agree with the reduced KdV predictions within this corridor. We regard dispersive pulses as a regime-specific complement to roll-wave dynamics, offering a condition-dependent contribution to mobility on gentle reaches rather than a universal explanation for long runout.
\end{abstract}

\maketitle

\section{Introduction}

Debris flows are high-velocity mixtures of water, soil, and rock capable of damaging infrastructure and reshaping landscapes. Their strongly nonlinear behavior arises from non-Newtonian rheology (e.g., yield stress, nonlinear viscosity), frictional effects, and grain–fluid interactions \cite{Iverson1997,Major1999}. These processes produce abrupt wave pulses and other features not captured by simple Newtonian shallow-water models~\cite{iverson1997debris,Aranson2006,Delannay2017}.

Steep-fronted surge trains in debris flows are classically described by roll-wave models that neglect dispersion and allow large amplitudes, beginning with Dressler~\cite{Dressler1949} and extended in related depth-averaged formulations, including boundary-layer regularizations~\cite{Balmforth2004}. These models successfully capture shock-like waves on steeper slopes.

Field and laboratory studies also show that gravity-driven mass flows can organize into coherent wave structures, including shock-like roll waves on steep slopes and weaker, more sinusoidal dispersive pulses on gentler slopes~\cite{plumerault2010high,Balmforth2004}. In the roll-wave regime, nonlinear steepening is balanced primarily by basal drag, whereas in the dispersive-pulse regime curvature-related normal stresses can balance weak nonlinearity. Which regime emerges depends on slope, the local Froude number, and effective dissipation.

We focus on gentle-slope reaches where fines-rich, low-resistance tails can sustain small-amplitude, long-wavelength pulses with a base-flow Froude number near unity. In this regime, a weakly nonlinear balance with curvature-related dispersion is admissible, and a Korteweg--de Vries (KdV) reduction applies. These pulses complement roll-wave surges and can contribute to momentum transport on gentle reaches, but they are not a universal explanation for long runout. In what follows we emphasize the long-wave small-$k$ setting in which the dispersive closure is intended to apply.

This paper proceeds as follows. We derive a depth-averaged model for gravity-driven mass flows and its KdV reduction, with effective coefficients that depend on frictional (Coulomb) and viscous–plastic parameters. We compile field constraints and map events onto a Froude–slope diagram to delineate roll-wave versus dispersive-pulse regime domains. We then validate the KdV description against numerical solutions of the full depth-averaged system. Definitions and derivations are presented in the Theory section, and notation is summarized in § Notation and base state and Appendix. For cross-referencing: Theory is §~\ref{sec:theory}; the list of symbols is in Appendix~\ref{sec:nomenclature}.

\section{Theory}
\label{sec:theory}

\subsection{Notation and base state}
\label{sec:notation}

We use a one-dimensional downstream coordinate $x$ along the channel and the flow time $t$. The flow depth is $h(x,t)$ and the depth-averaged streamwise velocity is $u(x,t)$. The uniform base state is $(h_0,u_0,\theta)$, where $h_0$ and $u_0$ are constants and $\theta$ is the bed angle. The bed slope is
$S = \tan\theta$.
Gravitational acceleration is $g$, and $\rho$ is the bulk density. For geometric summaries (e.g., distal-work budgets) we use $S=\tan\theta$, whereas in the momentum equation below we retain $g h \sin\theta$ explicitly so that the Coulomb base balance is exact; on gentle slopes $\sin\theta \approx S$ and $\cos\theta \approx 1$.

Perturbations about the base state are defined for the flow depth and velocity as follows:
\begin{equation}
\zeta(x,t) = h(x,t) - h_0,
\end{equation}
\begin{equation}
u'(x,t) = u(x,t) - u_0.
\end{equation}
The dimensionless free-surface perturbation is
\begin{equation}
\eta = \frac{\zeta}{h_0}.
\end{equation}
The linear shallow-water (long-wave) speed on the base state $(h_0,u_0)$ is
\begin{equation}
c_0 = \sqrt{g h_0}.
\end{equation}
When convenient, we scale the velocity perturbation by the base-state long-wave speed $c_0$ and write the dimensionless velocity perturbation
\begin{equation}
\hat u = \frac{u'}{c_0}.
\end{equation}
The event-scale Froude number, based on the depth-averaged speed $u$ and instantaneous depth $h$, is
\begin{equation}
\mathrm{Fr} = \frac{u}{\sqrt{g h}},
\end{equation}
and the base-flow Froude number is
\begin{equation}
\mathrm{Fr}_0 = \frac{u_0}{\sqrt{g h_0}}.
\end{equation}
Weak nonlinearity and long waves are measured by the amplitude ratio
\begin{equation}
\epsilon = \frac{a}{h_0},
\end{equation}
where $a$ is a characteristic free-surface elevation amplitude (e.g., a soliton amplitude or the crest elevation of a cnoidal wave) and $a \ll h_0$. The long-wave condition is
\begin{equation}
k h_0 \ll 1,
\end{equation}
where $k$ is the wavenumber and $\lambda = 2\pi/k$ is the periodic wavelength for cnoidal solutions. For solitary pulses, an analogous width $L_s$ is used; both are represented by the horizontal wave scale $L_{\mathrm{w}} \gg h_0$. The geometric reach length of the modeled channel used in distal-work budgets is $L$ and the channel width is $W_b$; these are distinct from the wave scale $L_{\mathrm{w}}$ (or $\lambda$).

In the depth-averaged formulation, rheology enters through basal and internal stresses. The internal (dispersive) component is modeled explicitly as follows.
Internal (dispersive) normal stress is denoted $\tau_{xx}$. In the long-wave closure used here it is modeled as
\begin{equation}
\tau_{xx} = \gamma \rho g h_0^{2} \zeta_{xx},
\label{eq:curvature_closure}
\end{equation}
where $\gamma$ is a dimensionless curvature–stiffness factor. For brevity, we refer to Eq.~(\ref{eq:curvature_closure}) as the curvature closure and cite it at subsequent uses rather than re-stating the model. This closure is intended for the small-$k$ long-wave regime ($k h_0 \ll 1$) and does not regularize high-$k$ behavior.

Basal shear stress is denoted $\tau_b$. We use two standard closures: a viscous–plastic (Bingham) form (with depth-averaged effective viscosity $\mu_{\mathrm{eff}}=\eta_{\mathrm{3D}}/h$) and a Coulomb form. Details, notation, and assumptions for both closures are given in § Rheological Equations (Sec.~\ref{sec:rheological_equations}). For distal-work diagnostics we use the tail-averaged friction $\mu_{\mathrm{tail}}$ where Eq.~(\ref{eq:work}) is introduced; we do not repeat formulas here.

We first collect the canonical shallow-water coefficients and governing relations for small-amplitude, long-wave free-surface gravity waves about a uniform base state. In the KdV model, the quadratic nonlinearity coefficient $\alpha_0$ and the dispersive coefficient $\beta_0$ are
\begin{equation}
\alpha_0 = \frac{3 c_0}{2 h_0}, \qquad \beta_0 = \frac{c_0 h_0^{2}}{6}.
\end{equation}
When depth-averaged rheology and the curvature closure \eqref{eq:curvature_closure} are included for gravity-driven mass flows, the coefficients are renormalized to $\alpha_{\mathrm{eff}}$ and $\beta_{\mathrm{eff}}$, with
\begin{equation}
\label{eq:16}
\beta_{\mathrm{eff}} = \beta_0\,\gamma = \frac{c_0 h_0^{2}}{6}\,\gamma.
\end{equation}
For a standard solitary wave (elevation) the KdV solution is
\begin{equation}
\eta(x,t) = A \sech^{2}\left(\frac{x - V t}{L_s}\right),
\end{equation}
where $A$ is the amplitude (for field surges, one may take $A=h_{1s}-h_0$, with $h_{1s}$ the peak flow depth of the first surge at the observation site), $L_s$ is the soliton width (when the standard KdV symbol $\Delta$ appears elsewhere, interpret $\Delta \equiv L_s$), and $V$ is the crest speed in the laboratory frame. For periodic solutions we use the Jacobi elliptic cosine $\mathrm{cn}(\cdot\mid m)$ with modulus $m\in(0,1)$; $K(m)$ and $E(m)$ are the complete elliptic integrals of the first and second kinds. The cnoidal wavenumber is $\kappa$ and the wavelength is
\begin{equation}
\Lambda = \frac{2K(m)}{\kappa}.
\end{equation}
(Details are summarized in §~\ref{sec:cnoidal_soliton_limit}.)

Symbols used in distal-work diagnostics include the tail depth $h_{\mathrm{tail}}$, the resisting stress $\tau_r$, the safety factor $SF$, and the work magnitude $W$; their definitions are given where first introduced in the text. All spatial and temporal derivatives are partial derivatives with respect to $x$ and $t$. A compact list of symbols is provided in Appendix~\ref{sec:nomenclature}.

\subsection{Debris-flow compilation}

\begin{figure*}[ht!]
\centering
\includegraphics[width=1.0\textwidth]{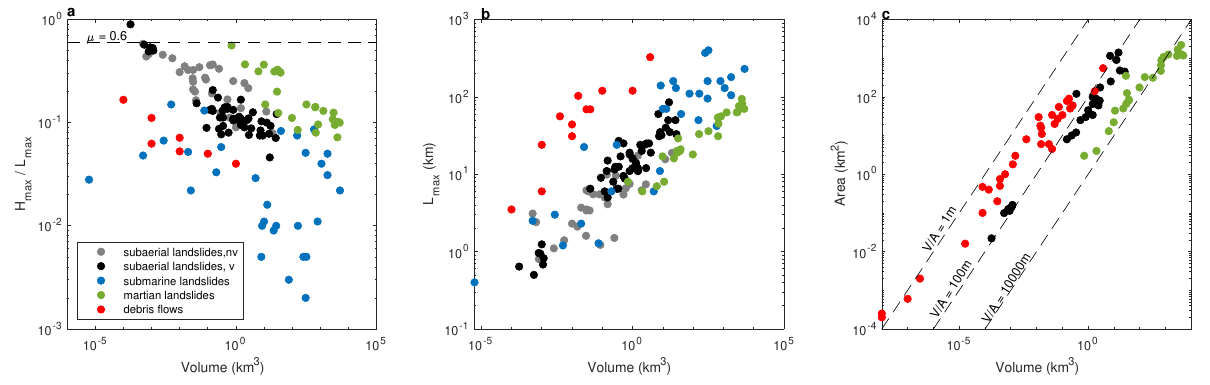}
\caption{Relationships among (a) apparent coefficient of friction ($H_{\mathrm{max}}/L_{\mathrm{max}}$), (b) runout distance ($L_{\mathrm{max}}$), (c) deposit area ($A$), and event volume ($V$) for 203 landslides and debris flows compiled by Legros~\cite{Legros2002}. Only debris-flow and subaerial cases are used in the quantitative analysis of Sec.~\ref{sec:soliton_energy}; submarine and planetary cases are shown for completeness. Symbols distinguish subaerial non-volcanic landslides (nv), subaerial volcanic landslides (v), submarine landslides, Martian landslides, and debris flows. Some entries lack estimates for one or more parameters. Each panel uses the subset of points with available data for the plotted quantities. The dataset is from Legros~\cite{Legros2002}, supplemented with values from Iverson~\cite{Iverson1997}.}
\label{fig:comp}
\end{figure*}

Legros compiled a data set of 203 long-runout landslides and debris flows that summarizes geometry and scaling across environments (Figure~\ref{fig:comp})~\cite{Legros2002}. The compilation spans subaerial volcanic and non-volcanic landslides, submarine landslides, Martian landslides, and debris flows from hillslope to megascale. We retain a trimmed set (debris-flow and subaerial entries) to provide a quantitative benchmark for the runouts considered here. The upper envelope in the $(H_{\max},L_{\max})$ plane motivates a distal energy-budget test presented later in the paper.

The compilation shows clear correlations among maximum runout distance $L_{\max}$, volume, and deposit area (Figure~\ref{fig:comp}). Legros interpreted those trends as evidence that spreading is governed chiefly by volume rather than by maximum fall height $H_{\max}$~\cite{Legros2002}. Submarine landslides and debris flows systematically outrun subaerial events of comparable volume, usually attributed to fluid support that lowers effective friction and hence the ratio $H_{\max}/L_{\max}$. Scatter about the trend reflects additional controls such as grain size, water versus ice content, slope geometry, and trigger type~\cite{Iverson1997,Legros2002}. What remains to be explained is how coherent wave pulses are sustained once the slope flattens; we evaluate whether weak dispersive pulses can help maintain momentum over distal, low-slope reaches.

To evaluate this possibility, we introduce two diagnostic quantities used throughout the paper: the safety factor, defined as the ratio between resisting stress $\tau_{r}$ and driving stress $\tau_{d}$ in the tail portion of the flow,
\begin{equation}
SF = \frac{\tau_{r}}{\tau_{d}},
\end{equation}
and the distal work magnitude $W$. From depth-averaged balances, these are
\begin{align}
\tau_{d} &= \rho g h_{\mathrm{tail}} S, 
& \tau_{r} &= \mu_{\mathrm{tail}} \rho g h_{\mathrm{tail}}, \nonumber\\
SF &= \frac{\tau_{r}}{\tau_{d}}, 
& W &= \lvert \tau_{d}-\tau_{r} \rvert W_{b} L^{2}, 
\label{eq:work}
\end{align}
where $h_{\mathrm{tail}}$ is the mean tail thickness (flow depth), $S$ is the mean bed slope, $L$ is the reach length, $W_{b}$ is the channel width, and $\mu_{\mathrm{tail}}$ is the tail-averaged Coulomb basal friction coefficient. We plot $\lvert W\rvert$ because on very gentle slopes $\tau_{r}$ can exceed $\tau_{d}$ for realistic $\mu_{\mathrm{tail}}$.
Equivalently,
\begin{equation}
W = \rho g h_{\mathrm{tail}} W_{b} L^{2} \lvert S - \mu_{\mathrm{tail}} \rvert.
\end{equation}
Here $S$ denotes the bed slope (approximately $\tan\theta$), and in the parameter sweep, we take $\mu_{\mathrm{tail}}\equiv\mu_s$.
On gentle slopes, $\sin\theta \approx \tan\theta$ and $\cos\theta \approx 1$. Under these approximations, the driving stress becomes $\tau_d \approx \rho g h_{\mathrm{tail}} S$, while the resisting stress is $\tau_r \approx \mu_{\mathrm{tail}} \rho g h_{\mathrm{tail}}$. The net stress is therefore $\tau_d - \tau_r \approx \rho g h_{\mathrm{tail}} (S - \mu_{\mathrm{tail}})$.
This is force $\times$ distance: $(\rho g h_{\mathrm{tail}} \lvert S - \mu_{\mathrm{tail}} \rvert)$ [Pa] $\times$ $(W_{b} L)$ [m$^{2}$] $\times$ $L$ [m] $=$ J. Thus, for fixed $h_{\mathrm{tail}}$, $W_{b}$, $S$, and $\mu_{\mathrm{tail}}$, one has $W \propto L^{2}$.

Equation~(\ref{eq:work}) is a bulk, path‐independent estimate that assumes slowly varying tail properties; in reality $h$ and $\mu_{\mathrm{tail}}$ are modulated by the wave train. We therefore use $|W|$ as a conservative diagnostic for distal energetics (upper bound when the tail thins along the reach, lower bound when it thickens), not as an exact path integral.

\begin{figure*}[ht!]
\centering
\includegraphics[width=0.65\textwidth]{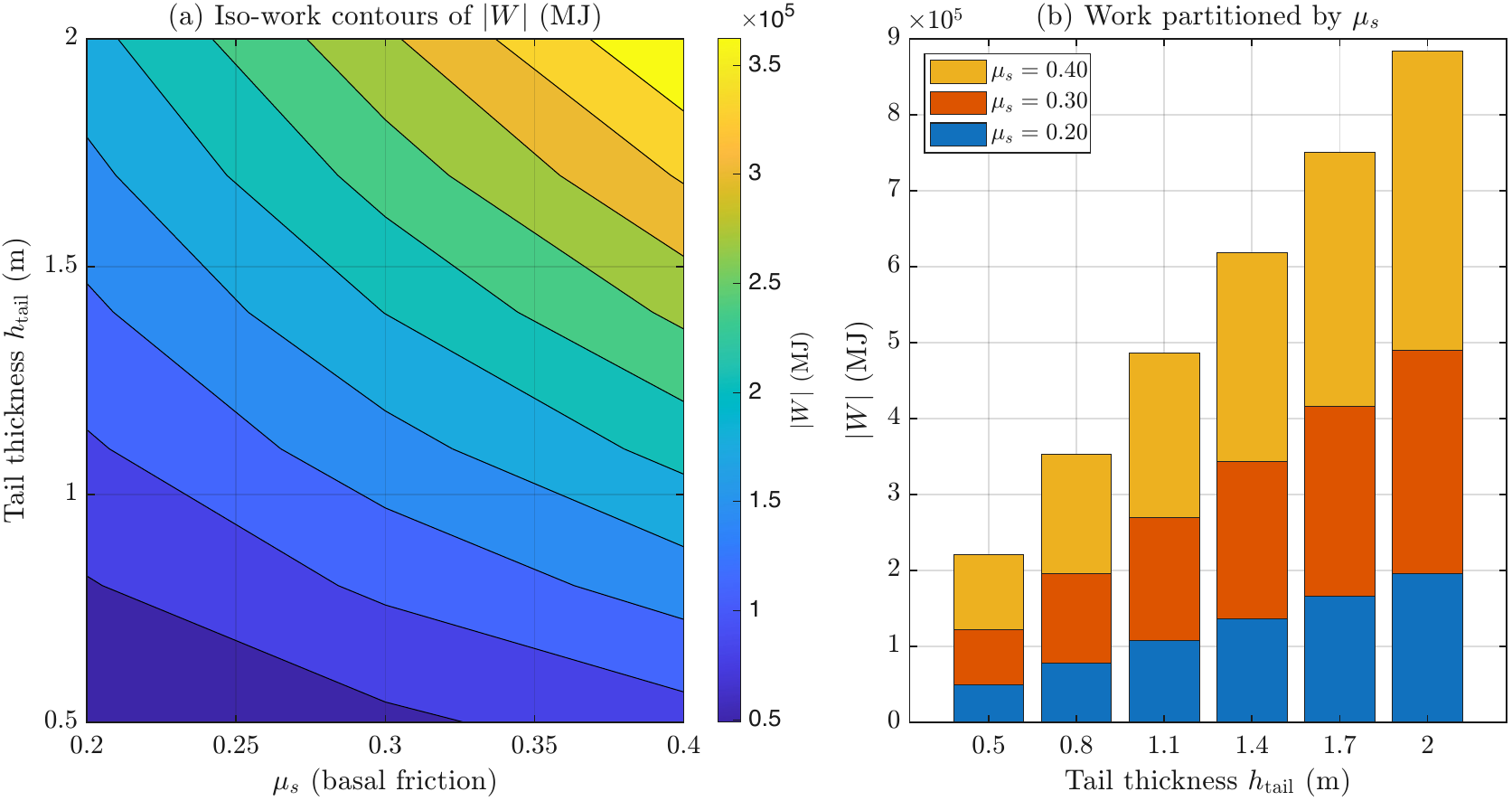}
\caption{
(a) Iso-work contours of the distal work magnitude $|W|$; (b) the same budget partitioned by the tail-averaged basal Coulomb coefficient $\mu_{\mathrm{tail}}$. Both panels use Eq.~(\ref{eq:work}) with $W=\rho g h_{\mathrm{tail}} W_b L^{2}\,|S-\mu_{\mathrm{tail}}|$ and the parameter grid described in the text. Take-home: $|W|$ scales linearly with $h_{\mathrm{tail}}$ and quadratically with $L$; near $S\approx\mu_{\mathrm{tail}}$, small changes in $\mu_{\mathrm{tail}}$ cause large relative changes in $|W|$.
}
\label{fig:energy}
\end{figure*}

For the parameter sweep supporting Figure~\ref{fig:energy}, we set $\rho=2000~\mathrm{kg\,m^{-3}}$ and $g=9.81~\mathrm{m\,s^{-2}}$; varied $h_{\mathrm{tail}}$ over $0.5$--$2.0~\mathrm{m}$ in $0.3~\mathrm{m}$ steps; and explored $\mu_s$ in the range $0.2$--$0.4$ (with lower effective values tested for sensitivity). Using event-specific inputs ($S=0.003$ for all three cases) and the paired reach--width values
\begin{equation}
(L,W_b)\in\{(1350~\mathrm{m}, 7~\mathrm{m}),\ (1600~\mathrm{m}, 9~\mathrm{m}),\ (1800~\mathrm{m}, 12~\mathrm{m})\},
\end{equation}
these choices generate a grid of $(SF,W)$ values that populate the contour and stacked--bar panels of Figure~\ref{fig:energy}. The work is evaluated using Eq.~(\ref{eq:work}),
\begin{equation}
W = \rho g h_{\mathrm{tail}} W_b L^{2}|S-\mu_s|,
\end{equation}
which scales linearly with $h_{\mathrm{tail}}$ and quadratically with $L$, and provides the starting point for the subsequent KdV analysis. Figure~\ref{fig:energy} highlights why long, low-slope reaches (large $L$, small $S$) dominate the distal work: $|W|\propto L^{2}$, and the budget is most sensitive when $S\approx \mu_{\mathrm{tail}}$.

Because the distal work depends on both material and geometric properties, we briefly summarize the rheological parameters that enter it. Debris flows exhibit complicated rheological behavior governed by solid volume fraction, grain-size distribution, and sediment composition \cite{OBrien1988,Iverson1997,Iverson2003,takahashi1978mechanical,Ancey2007}.
These factors set the bulk density $\rho$, yield stress $\tau_y$, and intrinsic (3D) dynamic viscosity $\eta_{\mathrm{3D}}$.
As indicated in Table~\ref{tab:rheological_properties}, field measurements, laboratory experiments, and model simulations indicate densities $\rho \approx 1600$--$2200\,\mathrm{kg\,m^{-3}}$, yield stresses $\tau_y \approx 10$--$4000\,\mathrm{Pa}$, and $\eta_{\mathrm{3D}} \approx 0.1$--$6000\,\mathrm{Pa\cdot s}$ \cite{Iverson2003}.

Increasing solids fraction tends to raise both yield strength and viscosity because of added particle contacts and frictional interactions~\cite{OBrien1988,Sosio2007}.
Grain-size distributions also matter~\cite{schippa2021yield,pellegrino2018laboratory}.
Field observations and flume experiments show that coarse grains concentrate at the flow head, which increases basal resistance; finer, water-rich material migrates rearward and creates a lower-resistance tail where small-amplitude dispersive pulses can propagate~\cite{Iverson1997,viroulet2018kinematics,meng2022formation}.

In the remainder of this paper, we compute $SF$ and $W$ for the Illgraben and Moscardo events, show that a train of weak solitary pulses can offset this work, and compare the resulting runouts with the upper envelope in Figure~\ref{fig:comp}, returning to the regime context in Sec.~\ref{sec:roll-vs-disp}.

\subsection{Wave-Like Phenomena in Gravity-Driven Mass Flows}

Field and laboratory studies show that gravity-driven mass flows can organize into coherent wave structures --- most often appearing as roll-wave trains on steeper slopes or as weaker, longer-wavelength pulses on gentler slopes --- that influence runout and velocity~\cite{Aranson2006,zanuttigh2007instability,Iverson2010,fei2023mu,arai2013occurrence}. Two idealized limits are often invoked in interpreting such patterns: shock-like roll waves, long studied in shallow-water theory, and weakly nonlinear dispersive waves describable by a KdV-type balance between nonlinear steepening and surface-curvature dispersion~\cite{Dressler1949,Balmforth2004}. Experimental studies on dense granular flows down rough inclines by Forterre \& Pouliquen~\cite{forterre2003long} documented long-surface-wave instabilities whose threshold and dispersion relations are well captured by depth-averaged Saint--Venant-type models, reinforcing the plausibility of dispersive behavior in non-Newtonian, granular mass flows. The wave regime realized in a given reach depends primarily on slope, local Froude number, and effective dissipation: roll-wave dynamics tend to dominate at high slopes and inertial forcing, while dispersive-pulse dynamics become more relevant on moderate slopes where dispersive effects are comparable to nonlinear steepening.

Additional evidence comes from controlled impulse-wave experiments and observations of localized solitary pulses in related geophysical and granular systems~\cite{Adeyemo2022,Fritz2002,sriram2016tsunami,Dominguez2019,Pelinovsky2006,Lynett2010}, as well as recent seismic monitoring that can resolve coherent wave signatures in natural debris flows~\cite{walter2023seismic,zhang2021analyzing}. Laboratory studies in other shallow-flow contexts also report clear regime transitions. For example, Plumerault \textit{et al.}~\cite{plumerault2010high} documented a ``pulse-waves'' regime at lower Reynolds numbers coexisting with a ``roll-waves'' regime at higher values, while Forterre \& Pouliquen~\cite{forterre2003long} identified a long-wave instability in dense granular flows with a well-defined threshold and dispersion relation, both consistent with depth-averaged stability theory. Together, these results support a regime view in which shock-dominated roll waves and dispersive pulses occupy different regions of parameter space and may coexist along a single event’s path.

In this study, we: (i) derive a depth-averaged, weakly nonlinear long-wave reduction leading to a KdV-type model with rheology-dependent coefficients; (ii) compile field constraints and place events on a Froude–slope diagram to relate roll-wave and dispersive-pulse regimes; and (iii) validate the reduced description against numerical solutions of the full rheological system. With this context, we next set down the governing depth-averaged equations that support these wave processes.

\begin{table*}[ht!]
\centering
\caption{Rheological parameters reported in the cited sources (intrinsic, 3D).}
\begin{tabular}{lll}
\toprule
Location & Rheological parameters (intrinsic 3D) & Ref. \\
\midrule
Jiangjia Ravine, China 
& $\eta_B = 0.4\text{--}15~\mathrm{Pa\cdot s}$ & \cite{Cui2005} \\
& $\tau_B = 10\text{--}300~\mathrm{Pa}$ & \cite{Cui2005} \\[5pt]

Wrightwood Canyon, USA 
& $\rho = 1620\text{--}2130~\mathrm{kg\,m^{-3}}$ & \cite{MortonCampbell} \\
& $\eta_B = 40\text{--}100~\mathrm{Pa\cdot s}$ & \cite{MortonCampbell} \\
& $\eta_N = 10\text{--}6000~\mathrm{Pa\cdot s}$ & \cite{MortonCampbell} \\
& $\eta_N = 210\text{--}600~\mathrm{Pa\cdot s}$ & \cite{SharpNobles1953} \\[5pt]

Miryang debris flow, South Korea 
& $\rho = 2000~\mathrm{kg\,m^{-3}}$ & \cite{jeong2024determining} \\
& $\eta_{B,L} = 0.8\text{--}6.2~\mathrm{Pa\cdot s}$ & \cite{jeong2024determining} \\
& $\tau_{B,L} = 63\text{--}2200~\mathrm{Pa}$ & \cite{jeong2024determining} \\[5pt]

Rossiga Valley, Italy 
& $\eta_B = 108\text{--}135~\mathrm{Pa\cdot s}$ & \cite{Sosio2007} \\
& $\tau_B = 3800\text{--}4200~\mathrm{Pa}$ & \cite{Sosio2007} \\
& $\eta_{B,L} = 0.6\text{--}27.9~\mathrm{Pa\cdot s}$ & \cite{Sosio2007} \\
& $\tau_{B,L} = 3.5\text{--}577~\mathrm{Pa}$ & \cite{Sosio2007} \\[5pt]

\bottomrule
\end{tabular}

\vspace{0.2cm}
\textit{Notes:} $\tau_B$ = Bingham yield strength; $\eta_B$ = Bingham viscosity; $\eta_N$ = Newtonian viscosity. Units are standardized to SI and reported as intrinsic (3D); ranges use en dashes. The depth-averaged effective viscosity is defined as $\mu_{\mathrm{eff}}=\eta_{\mathrm{3D}}/h$ (units Pa$\cdot$s/m) and is distinct from the dimensionless Coulomb coefficient $\mu_s$. The subscript $L$ denotes measurements on field-collected material tested in the laboratory (see cited sources).
\label{tab:rheological_properties}
\end{table*}

\subsection{Rheological Equations}
\label{sec:rheological_equations}

Debris flows often depart from the ideal shallow-water description and exhibit yield stress and nonlinear viscosity. We therefore use depth-averaged balances in which rheology enters through basal and internal stresses~\cite{Iverson1997,Major1999,OBrien1993,Ancey2007}.

Mass conservation is given by
\begin{equation}
\frac{\partial h}{\partial t} + \frac{\partial}{\partial x}(h u) = 0,
\end{equation}
where $h(x,t)$ is the local flow depth and $u(x,t)$ denotes the depth-averaged horizontal velocity.

Under a hydrostatic pressure approximation and a nearly uniform base state, the depth-averaged streamwise momentum balance with explicit slope forcing is
\begin{equation}
\label{eq:momentum_slope}
\frac{\partial (h u)}{\partial t}
+ \frac{\partial}{\partial x} \left(h u^{2} + \frac{g h^{2}}{2}\right)
= g h \sin\theta - \frac{h}{\rho}\frac{\partial \tau_{xx}}{\partial x} - \frac{\tau_b}{\rho},
\end{equation}
where $g$ is gravitational acceleration, $\rho$ is the bulk density, $\tau_{xx}$ is an internal (dispersive) normal stress, $\tau_b$ is the basal shear stress, and $S=\tan\theta$ is the geometric bed slope (used elsewhere for distal-work diagnostics). The uniform base state $(h_0,u_0,\theta)$ satisfies the steady balance
\begin{equation}
\label{eq:base_balance}
\rho g h_0 \sin\theta = \tau_b(h_0,u_0).
\end{equation}
Using the definitions given in §~\ref{sec:notation}, we write $h=h_0+\zeta$ and $u=u_0+u'$ and linearize \eqref{eq:momentum_slope} about $(h_0,u_0)$. The explicit slope term $g h \sin\theta$ then cancels with the basal contribution via \eqref{eq:base_balance} at leading order. The perturbation dynamics therefore omit an explicit slope forcing. A small mismatch between slope and basal resistance (for example $\sin\theta-\mu_s\cos\theta$, which reduces to $S-\mu_s$ when $\cos\theta\simeq 1$) produces only a weak KdV–Burgers-type diffusive correction at higher order.

In a full three-dimensional description, a Bingham plastic obeys
\begin{equation}
\label{eq:tau_bingham}
\tau = \tau_y + \eta_{\mathrm{3D}}\frac{\partial u}{\partial y},
\end{equation}
where $\tau$ is shear stress, $\tau_y$ is yield stress, $\eta_{\mathrm{3D}}$ is the intrinsic dynamic viscosity (Pa$\cdot$s), and $\partial u/\partial y$ is the shear rate. In a depth-averaged closure we approximate
\begin{equation}
\frac{\partial u}{\partial y} \approx \frac{u}{h},
\end{equation}
which motivates the effective basal relation
\begin{equation}
\tau_b \approx \tau_y + \mu_{\mathrm{eff}} u, \qquad
\mu_{\mathrm{eff}} = \frac{\eta_{\mathrm{3D}}}{h},
\end{equation}
with symbols summarized in Appendix~\ref{sec:nomenclature}. Motion initiates when the applied shear exceeds $\tau_y$; once mobilized, this provides a simple viscous–plastic law for the basal resistance.

For the Coulomb basal option (cf.\ §~\ref{sec:coulomb_main}),
\begin{equation}
\label{eq:tau_coulomb}
\tau_b = \mu_s \rho g h \cos\theta \operatorname{sgn}(u),
\end{equation}
where $\mu_s$ is a dimensionless basal friction coefficient. For distal-work diagnostics we use the tail-averaged coefficient $\mu_{\mathrm{tail}}$ where Eq.~(\ref{eq:work}) is introduced; detailed expressions are not repeated here.

The flux term $\partial_x(h u^{2})$ introduces a nonlinear interaction between $h$ and $u$, and the term $\partial_x(g h^{2}/2)$ contributes the hydrostatic pressure gradient. Together with basal and internal stresses, these terms set the balance between nonlinear steepening and dispersive spreading. Although $\tau_y$ may not appear explicitly in the leading-order wave equation once the flow is mobilized, it controls onset and therefore influences which waveforms can arise. The influence of $\tau_y$, $\mu_{\mathrm{eff}}$, $\mu_s$, and $h_0$ on admissible solitary and cnoidal regimes is quantified later (Fig.~\ref{fig:energy}; Secs.~\ref{sec:DebrisFlowAnalysis}, \ref{sec:fidelity}); here we note only that these parameters modulate the weak-nonlinear, long-wave balance~\cite{Coussot1997}.

\subsubsection{Frictional basal law at leading order}
\label{sec:coulomb_main}

Field-scale debris flows are commonly friction-dominated: basal shear is approximately proportional to normal pressure, and laboratory yield-stress measurements are often small compared to frictional stresses under natural overburden~\cite{Iverson2010}. Accordingly, we use the Coulomb basal law given in Eq.~(\ref{eq:tau_coulomb}), where $\mu_s$ is a dimensionless basal friction coefficient. Combining Eq.~(\ref{eq:tau_coulomb}) with the base-state balance in Eq.~(\ref{eq:base_balance}) yields $\tan\theta=\mu_s$, hence $S=\mu_s$.

Using the notation of §~\ref{sec:notation} and the slope–basal cancellation established by  \eqref{eq:momentum_slope} and \eqref{eq:base_balance}, the Coulomb term contributes no $\mathcal{O}(\epsilon)$ forcing to the weakly nonlinear, long-wave perturbation dynamics about a sign-definite base flow. Eliminating the velocity perturbation by the linear Saint–Venant slaving then yields the same KdV core as in the viscous–plastic variant,
\begin{equation}
\eta_t + (u_0+c_0) \eta_x + \alpha_{\mathrm{eff}} \eta \eta_x + \beta_{\mathrm{eff}} \eta_{xxx}=0,
\label{eq:kdv_coulomb_core}
\end{equation}
with $c_0$ as defined in §~\ref{sec:notation}, $\alpha_{\mathrm{eff}}\approx \alpha_0$, and $\beta_{\mathrm{eff}}=\beta_0 \gamma$, where $\gamma>0$ is the curvature–stiffness factor from the curvature closure in Eq.~(\ref{eq:curvature_closure}).

Small departures from exact base balance or slow spatial variability (for example, a small mismatch $\sin\theta-\mu_s\cos\theta$, reducing to $S-\mu_s$ when $\cos\theta\simeq 1$, or gentle variations of $h$ and $\mu_s$) produce a standard weak dissipative correction. At the reduced level, this appears as a KdV–Burgers term,
\begin{multline}
\eta_t + (u_0+c_0) \eta_x + \alpha_{\mathrm{eff}} \eta \eta_x \\
+ \beta_{\mathrm{eff}} \eta_{xxx}
= \delta \eta_{xx} + \text{higher-order terms},
\quad 0<\delta\ll 1,
\label{eq:kdv_burgers}
\end{multline}
where $\delta$ is an effective small diffusivity summarizing the near-balance mismatch and gentle inhomogeneities. In our simulations, we include basal work directly in the full-order equations; at the reduced level, we do not introduce ad hoc damping beyond noting that such a weak diffusive term is expected outside perfect balance.

The dominance of friction at field scale can be summarized by a yield number
\begin{equation}
\mathrm{Yu} \equiv \frac{\tau_y}{\rho g h_0 \sin\theta},
\end{equation}
which is typically $\mathrm{Yu}\ll 1$ for natural events (meters-thick flows and $S=\mathcal{O}(10^{-2}\text{–}10^{-1})$), implying that Bingham/Herschel–Bulkley parameters chiefly affect onset or weak damping rather than the leading-order KdV balance. This rationale explains why the KdV core~\eqref{eq:kdv_coulomb_core} is unchanged when the basal resistance is frictional.

\subsubsection{KdV Equation and Its Wave Solutions}

Under the ordering summarized in § Non-Dimensionalization and Scaling Analysis (weak nonlinearity and long waves), with $L_{\mathrm{w}}$ the wave scale distinct from the reach length $L$ used in Eq.~(\ref{eq:work}), we expand about a uniform base state and introduce slow variables that capture the modulation of right-running waves. Write
\begin{equation}
h(x,t) = h_0 + \epsilon \eta(x,t), 
\quad u(x,t) = u_0 + \mathcal{O}(\epsilon),
\end{equation}
Equation \eqref{eq:kdv} follows from the depth-averaged mass and momentum balances (see Theory) under the ordering in § Non-Dimensionalization and Scaling Analysis, together with the curvature closure in Eq.~(\ref{eq:curvature_closure}); detailed multiple-scale steps are given in § Derivation of the KdV Equation from Depth-Averaged Equations.
Then, at leading order, the free-surface perturbation satisfies the KdV equation
\begin{equation}
\eta_t + c \eta_x + \alpha \eta \eta_x + \beta \eta_{xxx}=0,
\label{eq:kdv}
\end{equation}
where $c$ is the linear convective speed on the base state (laboratory frame $c=u_0+c_0$, or in a frame translating with the base flow $c=c_0$, with $c_0$ as defined in §~\ref{sec:notation}), and $\alpha$ and $\beta$ are the effective coefficients governing weak nonlinearity and dispersion, respectively.

In the Newtonian shallow-water limit (no yield stress and no internal dispersion beyond hydrostatic pressure) with no background shear, we use the canonical coefficients $c_0$, $\alpha_0$, and $\beta_0$ as defined in §~\ref{sec:notation}.
With a uniform base current $u_0$, one may either write a convected KdV by replacing $\partial_t$ with $\partial_t + u_0 \partial_x$, or apply a Galilean change of variables $x \mapsto x - u_0 t$, in which case the coefficients above remain unchanged.

For the depth-averaged momentum equation \eqref{eq:momentum_slope}, alternative third-order, layer-averaged dispersive closures (e.g., Boussinesq/Serre–Green–Naghdi–type) reduce, under shallow-slope/long-wave assumptions, to the same leading-order KdV dispersion (small-$k$ expansion of $c_{\mathrm{ph}}(k)$) as obtained with the curvature closure \eqref{eq:curvature_closure}; hence using the curvature term for $\tau_{xx}$ is illustrative rather than restrictive~\cite{Balmforth2004}. Non-Newtonian effects—such as a yield stress $\tau_y$, a depth-averaged effective viscosity $\mu_{\mathrm{eff}}$, and curvature-related internal stresses—renormalize the KdV coefficients. To avoid overloading symbols, we reserve $\gamma$ for the curvature (dispersive) prefactor that appears in the internal normal-stress model (see Sec.~\ref{sec:dispersion}); the slow modulation in the multiple-scale expansion is encoded by the small parameter $\epsilon$ and the stretched variables, not by $\gamma$.

The curvature closure is intended for the long-wave branch, where $k h_{0}\ll 1$. Here $k$ is the spatial wavenumber (rad m$^{-1}$), so this condition is equivalent to a wavelength $\lambda=2\pi/k$ satisfying $\lambda \gg 2\pi h_{0}$. We impose $\gamma>0$, which gives $\beta_{\mathrm{eff}}>0$ and the small-$k$ phase speed in the laboratory frame
$c_{\mathrm{ph}}(k)=u_{0}+c_{0}-\beta_{\mathrm{eff}}k^{2}+\mathcal{O}(k^{4})$ (equivalently, in a frame translating with $u_0$, $c_{\mathrm{ph,rel}}(k)=c_{0}-\beta_{\mathrm{eff}}k^{2}+\mathcal{O}(k^{4})$) that decreases with $k$; the long-wave branch is then well posed in this limit. At shorter waves, third-order truncations can show spurious high-$k$ behavior; alternative regularizations (e.g., Boussinesq/Serre--Green--Naghdi or boundary-layer formulations as in Balmforth \& Mandre~\cite{Balmforth2004}) preserve the same small-$k$ limit (i.e., reproduce the same leading-order dispersive coefficient $\beta_{\mathrm{eff}}$ in Eq.~(\ref{eq:16})) and can improve short-wave spectra when their coefficient constraints are met. All simulations are configured so that the resolved spectra remain in the $k h_{0}\ll 1$ window where the closure is asymptotically valid.

Accordingly, we write
\begin{equation}
\label{eq:35}
\alpha= \alpha_{\mathrm{eff}}(h_0,u_0; \tau_y, \mu_{\mathrm{eff}},\mu_s),
\quad
\beta=\beta_{\mathrm{eff}}(h_0; \gamma),
\end{equation}
with $\gamma$ entering $\beta_{\mathrm{eff}}$ through the curvature (normal-stress) closure. Once the flow is mobilized, $\tau_y$ and $\mu_{\mathrm{eff}}$ chiefly renormalize $\alpha_{\mathrm{eff}}$; at KdV order $\beta_{\mathrm{eff}}$ depends on $\gamma$ (and $h_0$ via $c_0$) but not on $\tau_y$ or $\mu_{\mathrm{eff}}$, with any such influence deferred to higher-order corrections.

For the remainder of the asymptotic development, we use $L_{\mathrm{w}}$ exclusively for the horizontal wavelength or pulse width in the KdV scaling (typically $\gg h_0$), and reserve $L$ for the geometric reach length in the distal-work calculation of Eq.~(\ref{eq:work}). This separation avoids conflating the $L^2$ scaling in the energy budget with the $\mathcal{O}(L_{\mathrm{w}}^{-2})$ scaling in the KdV ordering.

In the co-moving (Galilean) coordinate $\xi = x - c t$ with $c=u_0+c_0$ (derivatives taken at fixed $\xi$), the free-surface perturbation satisfies
\begin{equation}
\eta_t + \alpha \eta \eta_\xi + \beta \eta_{\xi\xi\xi} = 0,
\end{equation}
where $\alpha$ and $\beta$ are the effective nonlinear and dispersive coefficients defined above. In the laboratory frame, the linear advection term $(u_0+c_0)\eta_x$ is restored. The quadratic term $\alpha \eta \eta_x$ represents nonlinear advection that steepens wave crests, while the cubic derivative term $\beta \eta_{xxx}$ represents dispersion that spreads them. When nonlinear steepening and dispersion are of the same asymptotic order, the KdV equation supports stable traveling-wave solutions, including solitary pulses and periodic cnoidal waves. For debris flows with small relative amplitude $(\epsilon \ll 1)$ and long wavelength $(L_{\mathrm{w}} \gg h_0)$, this reduced equation provides a tractable description of coherent free-surface structures. In the sections that follow, we determine $\alpha_{\mathrm{eff}}$ and $\beta_{\mathrm{eff}}$ from the chosen rheology and internal-stress closure, and verify that cnoidal and soliton solutions of the KdV model persist in simulations of the full depth-averaged equations.

\subsubsection{Cnoidal Waves and the Soliton Limit}
\label{sec:cnoidal_soliton_limit}

Traveling-wave solutions of \eqref{eq:kdv} arise by seeking profiles that are stationary in a moving frame, i.e., set $\eta(x,t)=\eta(\xi)$ with the traveling-wave coordinate $\xi \equiv x - V t$ (so $\partial_x=\partial_\xi$ and $\partial_t=-V \partial_\xi$). Integrating once yields a cubic potential in $\eta(\xi)$. Writing the three real roots as $\eta_1\le \eta_2\le \eta_3$, the periodic cnoidal family can be expressed as
\begin{equation}
\eta(\xi)= \eta_2 + \bigl(\eta_3-\eta_2\bigr)\mathrm{cn}^{2}\bigl(\kappa \xi \mid m\bigr).
\end{equation}
Here cn denotes the Jacobi elliptic cosine with modulus $m$; $K$ and $E$ denote the complete elliptic integrals of the first and second kinds, respectively. Standard references for these formulas are \cite{drazin1989solitons,whitham1974linear}.
The elliptic modulus $m$ and wavenumber $\kappa$ are
\begin{equation}
m=\frac{\eta_3-\eta_2}{\eta_3-\eta_1}, 
\quad
\kappa=\sqrt{\frac{\alpha (\eta_3-\eta_1)}{12 \beta}},
\end{equation}
and the phase speed satisfies
\begin{equation}
V = c + \frac{\alpha}{3} \bigl(\eta_1+\eta_2+\eta_3\bigr).
\end{equation}
The amplitude is $A=\eta_3-\eta_2$, the lower (minimum) level is $\eta_2$, and the wavelength is
\begin{equation}
\Lambda=\frac{2K(m)}{\kappa}.
\end{equation}
The mean over one wavelength is
\begin{equation}
\bar\eta=\eta_2+\bigl(\eta_3-\eta_2\bigr) \frac{E(m)}{K(m)}.
\end{equation}
As $m\to0$, $\mathrm{cn}(\cdot\mid m)\to\cos(\cdot)$ and the wave is nearly sinusoidal; as $m\to1^{-}$, $K(m)\to\infty$ so $\Lambda=2K(m)/\kappa\to\infty$ and the cnoid limits to the solitary pulse.

For completeness we record a minimal derivation of the solitary-pulse relations. In the traveling coordinate $\xi=x-Vt$, the KdV equation \eqref{eq:kdv} gives $-V \eta' + c \eta' + \alpha \eta \eta' + \beta \eta'''=0$. Integrating once with the solitary boundary conditions ($\eta,\eta'\to 0$ as $|\xi|\to\infty$) yields
\begin{equation}
\beta \eta'' = (V-c) \eta - \frac{\alpha}{2} \eta^{2}.
\end{equation}
The decaying solution is the $\sech^2$ pulse
\begin{equation}
\eta(\xi)=A \sech^{2} \left(\frac{\xi}{L_s}\right),
\end{equation}
which, upon substitution, enforces the relations
\begin{equation}
L_s^{2}=\frac{12 \beta}{\alpha A}, \quad V-c=\frac{\alpha A}{3}.
\end{equation}
In the soliton limit $(m\to 1, \eta_1\to \eta_2)$, the profile reduces to the form above, with laboratory- and co-moving-frame speeds
\begin{equation}
V_{\mathrm{lab}}=u_0+c_0+\frac{\alpha A}{3}, \quad
V_{\mathrm{rel}}=c_0+\frac{\alpha A}{3},
\end{equation}
where $c_0$ is as defined in §~\ref{sec:notation}. Here $L_s$ denotes the soliton (pulse) width, defined by $L_s^{2}=12 \beta/(\alpha A)$. For fixed $\alpha$ and $\beta$, larger $|A|$ implies a narrower pulse ($L_s \propto |A|^{-1/2}$ via the preceding relation). A real width requires $(\alpha A)/\beta>0$; in particular, elevation solitons ($A>0$) require $\alpha\beta>0$, whereas depression solitons ($A<0$) require $\alpha\beta<0$.

Using the curvature closure summarized in Eq.~(\ref{eq:16}), the observed $(A,L_s,V)$ of an elevation soliton determine the dispersive stiffness via
\begin{equation}
\gamma=\frac{\alpha A L_s^{2}}{2 c_0 h_0^{2}}
=\frac{3 L_s^{2}\bigl(V-u_0-c_0\bigr)}{2 c_0 h_0^{2}}.
\end{equation}

\subsubsection{Stability Analysis of Soliton Solutions}

Linearizing \eqref{eq:kdv} about the solitary wave in the co-moving coordinate $\xi=x-Vt$ gives
\begin{equation}
w_t = -\partial_\xi \Bigl[(c-V)w + \alpha \eta_{\mathrm{s}}(\xi) w + \beta w_{\xi\xi}\Bigr],
\end{equation}
where $\eta_{\mathrm{s}}(\xi)=A \sech^2\bigl(\xi/L_s\bigr)$ and $w(\xi,t)$ is the perturbation. Here $L_s$ denotes the soliton (pulse) width; when the classical KdV symbol $\Delta$ appears, interpret $\Delta \equiv L_s$. The width is given by $L_s^{2}=12 \beta/(\alpha A)$.
This can be written in Hamiltonian form $w_t = J L w$ with $J=-\partial_\xi$ (skew-adjoint) and $L=\beta \partial_\xi^2+(c-V)+\alpha \eta_{\mathrm{s}}$ (self-adjoint). The spectrum has a simple eigenvalue at zero due to translation symmetry $(w\propto \eta_{\mathrm{s},\xi})$ and no spectrum with positive real part. Hence the KdV soliton is spectrally stable and, in fact, is orbitally stable in $H^1$ in the sense of Benjamin \cite{Benjamin1972}: small perturbations remain close (modulo phase) for all time. This property supports the use of solitary pulses as robust carriers of momentum and energy in weakly nonlinear, long-wave debris-flow regimes.

\subsubsection{Derivation of the KdV Equation from Depth-Averaged Equations}

We derive the reduced equation starting from the depth-averaged continuity and momentum balances stated in Theory. For completeness, we write the continuity equation
\begin{equation}
\frac{\partial h}{\partial t}
+\frac{\partial}{\partial x}\bigl(h u\bigr)
=0,
\label{eq:continuity2}
\end{equation}
and refer to the momentum equation in its explicit slope form as Eq.~(\ref{eq:momentum_slope}). The base-state balance Eq.~(\ref{eq:base_balance}) implies slope–basal cancellation at leading order in the perturbation. The hydrostatic term $\partial_x( g h^{2}/2 )$ supplies the leading pressure contribution.

We seek a KdV reduction under the weakly nonlinear, long-wave ordering summarized in § Non-Dimensionalization and Scaling Analysis and employ the curvature closure in Eq.~(\ref{eq:curvature_closure}). Linearization about $(h_0,u_0)$, elimination of the velocity perturbation via the right-running Saint–Venant slaving, and retention of the lowest-order nonlinear and dispersive terms then yield the KdV balance quoted in § KdV Equation and Its Wave Solutions.

\subsubsection{Non-Dimensionalization and Scaling Analysis}

To isolate the primary balance, let $L_{\mathrm{w}}$ be the characteristic wave horizontal scale and define
\begin{equation}
\tilde{x}= \frac{x}{L_{\mathrm{w}}}, \quad
\tilde{t}= \frac{c_0 t}{L_{\mathrm{w}}}.
\end{equation}
Here $c_0$ is as defined in §~\ref{sec:notation}; it is not redefined.
We reserve $L_{\mathrm{w}}$ for the wave scale used in the asymptotics; the geometric reach length $L$ appears only in the distal-work budget of Eq.~(\ref{eq:work}).
Then,
\begin{equation}
\frac{\partial}{\partial x}= \frac{1}{L_{\mathrm{w}}} \frac{\partial}{\partial \tilde{x}}, \quad
\frac{\partial}{\partial t} = \frac{c_0}{L_{\mathrm{w}}} \frac{\partial}{\partial \tilde{t}}.
\end{equation}
We adopt the weakly nonlinear, long-wave ordering $\epsilon\sim(h_0/L_{\mathrm{w}})^2$, which balances quadratic nonlinearity with third-order dispersion.
We expand the fields about a uniform base state $(h_0,u_0)$:
\begin{equation}
\label{eq:53}
h(x,t) = h_0 \bigl[1 + \epsilon \eta(\tilde{x}, \tilde{t})\bigr], 
\quad 
u(x,t) = u_0 + \epsilon c_0 \hat u(\tilde{x}, \tilde{t}),
\end{equation}
using $\hat u=u'/c_0$ from §~\ref{sec:notation} to avoid introducing a new variable $\delta u$. A Galilean shift $x\mapsto x-u_0 t$ removes uniform advection; equivalently one may keep $u_0$ and interpret $c$ as a convected wave speed. With the curvature closure in Eq.~(\ref{eq:curvature_closure}), the internal-stress gradient produces a third derivative of $\eta$ with coefficient scaling $\beta=\mathcal{O}(c_0 h_0^2)$, consistent with the ordering stated in this subsection, and matching the nonlinear term at order $\mathcal{O}(\epsilon)$.

Substituting the scaled fields into the governing balances (see §~\ref{sec:theory}) and retaining $\mathcal{O}(\epsilon)$ terms identifies the same leading mechanisms already summarized: nonlinear advection via $h u^2$, curvature-induced dispersion through $\tau_{xx}$, and basal resistance through the chosen closure. Consistent with Eq.~(\ref{eq:35}) (and the curvature closure in Eq.~(\ref{eq:curvature_closure})), we therefore take $\alpha$ and $\beta$ as in Eq.~(\ref{eq:35}) without repeating the intermediate algebra.

Higher-order corrections (e.g., fifth-order dispersion or quadratic nonlinearity) and three-dimensional effects are deferred to later sections, while the present focus is the $\mathcal{O}(\epsilon)$ KdV balance that admits cnoidal and solitary waves.

\subsubsection{Expansion of the Continuity and Momentum Equations}
\label{sec:expansion}

We use the small-amplitude ($\epsilon= a/h_0\ll1$) and long-wave ($L_{\mathrm{w}} \gg h_0$) scaling with the KdV ordering $\epsilon \sim(h_0/ L_{\mathrm{w}})^2$ so that weak nonlinearity and weak dispersion enter at the same order. Working from the dimensional forms of Eqs.~\eqref{eq:continuity2} and \eqref{eq:momentum_slope}, and using the base-state slope–basal cancellation established earlier, we expand each equation in powers of $\epsilon$ and retain $\mathcal{O}(\epsilon)$ terms.

Starting with
\begin{equation}
\frac{\partial h}{\partial t}+ \frac{\partial}{\partial x} \bigl(h u\bigr)=0,
\label{eq:continuity_original}
\end{equation}
we substitute
\begin{equation}
\label{eq:55}
h(x,t) = h_0 \bigl[1 + \epsilon \eta(x,t)\bigr],
\quad
u(x,t) = u_0 + \epsilon c_0 \hat u(x,t),
\end{equation}
which is the same small-amplitude expansion as Eq.~(\ref{eq:53}), here written without tildes for brevity (the order-one fields $\eta$ and $\hat u$ are those of the scaled variables).
Then
\begin{equation}
\frac{\partial h}{\partial t} = h_0 \epsilon \eta_t,
\quad
h u = h_0 u_0 \Bigl[1 + \epsilon \eta\Bigr] + h_0 \epsilon c_0 \hat u + \mathcal{O}(\epsilon^2),
\end{equation}
and hence
\begin{equation}
\frac{\partial (h u)}{\partial x}
= h_0 u_0 \epsilon \eta_x + h_0 \epsilon c_0 \hat u_x + \mathcal{O}(\epsilon^2).
\end{equation}
Substituting into \eqref{eq:continuity_original} gives
\begin{equation}
\eta_t + u_0 \eta_x + c_0 \hat u_x = \mathcal{O}(\epsilon).
\label{eq:scaled_continuity_final}
\end{equation}
At leading order we therefore have a single kinematic relation linking the free-surface and velocity perturbations.

Starting from the momentum balance in Eq.~(\ref{eq:momentum_slope}), substituting the small-amplitude expansions of Eq.~(\ref{eq:53}), and using the base-state balance Eq.~(\ref{eq:base_balance}) to cancel the explicit slope at leading order, the $\mathcal{O}(\epsilon)$ terms give
\begin{multline}
h_0 u_0 \epsilon \eta_t + h_0 \epsilon c_0 \hat u_t
+ h_0 u_0^2 \epsilon \eta_x + 2 h_0 \epsilon u_0 c_0 \hat u_x
+ g h_0^2 \epsilon \eta_x \\
= -\frac{h_0}{\rho} \partial_x \tau_{xx}^{(1)}
-\frac{\tau_b^{(1)}}{\rho}
+ \mathcal{O}(\epsilon^2).
\label{eq:momentum_expanded}
\end{multline}
Here $\tau_{xx}^{(1)}$ and $\tau_b^{(1)}$ denote $\mathcal{O}(\epsilon)$ perturbations. For mobilized flow, yield stress primarily affects onset; at KdV order we neglect $\tau_b^{(1)}$ and retain the curvature contribution via Eq.~(\ref{eq:curvature_closure}).

We now close the linear part. For long, shallow waves about a uniform base state $(h_0,u_0)$, the dimensional right-going Saint–Venant mode satisfies
\begin{equation}
u' \approx \frac{c_0}{h_0} \zeta,\quad
u'=u-u_0,\quad \zeta=h-h_0.
\end{equation}
Here $c_0$ is as defined in §~\ref{sec:notation}.
This corresponds to the linearization of the right-running Saint–Venant Riemann invariant $R^{+}=u+2\sqrt{g h}$ about $(h_0,u_0)$, yielding $u'-(c_0/h_0) \zeta=0$ along $dx/dt=u_0+c_0$.

In a co-moving (Galilean) frame, linearizing the right-running Saint–Venant invariant about $(h_0,u_0)$ gives the dimensional relation $u' \approx (c_0/h_0)\zeta$ (see §~\ref{sec:notation} for $u'$ and $\zeta$). Adopting the frame-robust scaling
\begin{equation}
\hat{u} \equiv \frac{u'}{c_0}, \quad \eta \equiv \frac{\zeta}{h_0},
\end{equation}
yields the slaving
\begin{equation}
\hat{u} \approx \eta,
\label{eq:slaving_relation}
\end{equation}
which we use henceforth. In particular, $\hat u$ replaces any auxiliary $\delta u$ variable throughout.

Substituting \eqref{eq:slaving_relation} into \eqref{eq:scaled_continuity_final} and \eqref{eq:momentum_expanded}, eliminating $\hat u$ and $\hat u_t$, and retaining the curvature contribution
\begin{equation}
-\frac{h}{\rho}\partial_x\tau_{xx}
\leadsto 
\beta \eta_{xxx}
\quad
\text{with}\quad
\beta=\mathcal{O}(c_0 h_0^2),
\end{equation}
one obtains a unidirectional evolution equation of KdV type for the free-surface perturbation,
\begin{equation}
\eta_t + c \eta_x + \alpha \eta \eta_x + \beta \eta_{xxx}=0,
\label{eq:kdv_recovered}
\end{equation}
where $c$ is the linear convected speed (equal to $u_0+c_0$ in the laboratory frame or $c_0$ in a frame moving with $u_0$), and the effective coefficients $\alpha$ and $\beta$ encode the weakly nonlinear and dispersive corrections implied by the rheology and by the chosen curvature closure for $\tau_{xx}$. The basal-drag and yield-stress terms first modify $\alpha$ at higher order or through weak damping; those effects are deferred to later sections.

The weakly nonlinear, long-wave expansion together with the linear slaving~\eqref{eq:slaving_relation} isolates the leading interplay between advective steepening and curvature-induced dispersion, yielding the KdV balance in~\eqref{eq:kdv_recovered}.

\subsubsection{Isolation of Dispersive Effects}
\label{sec:dispersion}

We represent curvature-induced normal stresses in terms of the dimensional free-surface perturbation $\zeta(x,t)=h(x,t)-h_0$ and write
\begin{equation}
\tau_{xx}=\gamma \rho g h_0^{2} \zeta_{xx},
\label{eq:69}
\end{equation}
with $\gamma$ a dimensionless curvature–stiffness parameter. This is the same curvature closure referenced elsewhere (cf.\ Eq.~(\ref{eq:curvature_closure})). Since $\zeta_{xx}$ has units $1/\mathrm{m}$, the product $\rho g h_0^{2}\zeta_{xx}$ has units of pressure; substituting this closure into the momentum balance supplies the third derivative needed for KdV-type dispersion.

Under the long-wave, multiple-scale expansion, eliminating the velocity perturbation via the right-running Saint–Venant slaving yields the leading dispersive coefficient as in Eq.~(\ref{eq:16}). Equivalently, linearizing the depth-averaged balances about $(h_0,u_0)$ with a plane-wave ansatz $\zeta,u' \propto e^{i(kx-\omega t)}$ and using the curvature closure (Eq.~(\ref{eq:curvature_closure})) gives the small-wavenumber dispersion relation
\begin{equation}
(\omega - k u_0)^2 = g h_0 k^2 \left[1 - \frac{\gamma}{3}(k h_0)^2\right] + \mathcal{O}\big((k h_0)^4\big).
\end{equation}
By “small-$k$ dispersion curve” we mean the laboratory-frame phase-speed expansion $c_{\mathrm{ph}}(k)=\omega/k$ for $k h_0\ll 1$, which here is
\begin{equation}
c_{\mathrm{ph}}(k)=u_0 + c_0 - \beta_{\mathrm{eff}} k^2 + \mathcal{O}(k^4),
\end{equation}
recovering the same leading-order $\beta_{\mathrm{eff}}$ as in Eq.~(\ref{eq:16}).
In what follows we take
\begin{equation}
\gamma>0,
\label{eq:polarity_choice}
\end{equation}
so that $\beta_{\mathrm{eff}}>0$; with $\alpha>0$ this admits elevation solitons ($\alpha\beta_{\mathrm{eff}}>0$). Taking $\gamma<0$ would simply reverse the dispersion sign and favor depression solitons.

The factor $\gamma$ sets the scale of the small-$k$ dispersion and is generally site dependent, so it is not uniquely determined by a single pulse. If a small-$k$ dispersion curve $c_{\mathrm{ph}}(k)$ is available,
\begin{equation}
\gamma \approx -\frac{6}{c_0 h_0^2} 
\left.\frac{\partial}{\partial(k^2)} \big[c_{\mathrm{ph}}(k)-(u_0+c_0)\big]\right|_{k\to 0}.
\end{equation}
Alternatively, for an elevation soliton one may infer $\gamma$ from the observed $(A,L_s,V)$ via the solitary-wave relation given earlier in §~\ref{sec:cnoidal_soliton_limit}, avoiding repetition here.

At shorter waves, third-order truncations can exhibit spurious high-$k$ behavior; alternative regularizations (e.g., Boussinesq/Serre–Green–Naghdi or boundary-layer regularizations as in Balmforth \& Mandre~\cite{Balmforth2004}) match the same long-wave coefficient $\beta_{\mathrm{eff}}$ of Eq.~(\ref{eq:16}) and can improve short-wave spectra when their coefficient constraints are satisfied. All simulations are configured so that resolved spectra remain in the $k h_0\ll 1$ window where this closure is asymptotically valid.

\subsubsection{Derivation of the Nonlinear Contribution}

The quadratic nonlinearity in the reduced equation arises from the advective fluxes in the depth-averaged Saint–Venant system, not from an imposed slow-time ansatz. In the weakly nonlinear, long-wave limit the expansion produces a term proportional to $\eta \eta_x$ with a coefficient set by the base depth and the linear wave speed.

Using the small-amplitude expansions of Eq.~(\ref{eq:55}) (with $\eta$ and $\hat u$ order-one fields in the scaled variables) and the notation of §~\ref{sec:notation}, the right-running Saint–Venant slaving gives $\hat u \approx \eta$ at leading order. To isolate intrinsic wave propagation, we evaluate the balance in a frame translating with the base flow so that the linear advection speed in the KdV term is $c\approx c_0$. Retaining terms through $\mathcal{O}(\epsilon)$ in the mass and momentum equations and eliminating $\hat u$ via the slaving yields the standard weakly nonlinear balance
\begin{equation}
\eta_t + c \eta_x + \alpha_0 \eta \eta_x = \text{dispersive and higher-order terms},
\end{equation}
where the leading quadratic coefficient is $\alpha_0$ as defined in §~\ref{sec:notation}.
Thus the steepening strength is set by hydrostatic nonlinearity and momentum advection, and scales like $c/h_0$.

Rheology can renormalize this coefficient without changing its scaling. We write
\begin{equation}
\alpha = \Phi_{\mathrm{nl}} \alpha_0,
\end{equation}
where $\Phi_{\mathrm{nl}}$ depends smoothly on parameters such as effective viscosity, yield or Coulomb friction, and base-flow Froude number. For the parameter ranges considered here, $\Phi_{\mathrm{nl}}$ remains near unity, so the admissibility and regime inferences based on $\alpha_0$ are unchanged to leading order.

\subsubsection{Final Formulation of the KdV Equation}
\label{sec:weak_damping}

Collecting the balances summarized in § Non-Dimensionalization and Scaling Analysis and §~\ref{sec:dispersion}, and using the slaving \eqref{eq:slaving_relation}, the KdV form obtained in Eq.~(\ref{eq:kdv_recovered}) applies with the effective coefficients of Eq.~(\ref{eq:35}).  In the co-moving frame,
\begin{equation}
\eta_t + \alpha_{\mathrm{eff}} \eta \eta_x + \beta_{\mathrm{eff}} \eta_{xxx} = 0,
\label{eq:kdv_final_comoving}
\end{equation}
and the laboratory-frame form restores uniform linear advection by the base current and gravity wave speed,
\begin{equation}
\eta_t + (u_0 + c_0) \eta_x + \alpha_{\mathrm{eff}} \eta \eta_x + \beta_{\mathrm{eff}} \eta_{xxx} = 0.
\label{eq:kdv_final_lab}
\end{equation}
The canonical coefficients $c_0$, $\alpha_0$, and $\beta_0$ are as defined in §~\ref{sec:notation}, and the renormalized values follow Eq.~(\ref{eq:35}) with the curvature closure Eq.~(\ref{eq:curvature_closure}) (cf.\ Eq.~(\ref{eq:16})). A Galilean shift does not alter $\alpha$ or $\beta$; it removes or restores $(u_0+c_0)\eta_x$ depending on the frame. Basal resistance depends on absolute slip, so frame shifts do not strictly preserve dissipation; departures from exact base balance appear as a weak KdV–Burgers correction (diffusive $\eta_{xx}$), small in the regime considered and handled in the full-order numerics. We adopt the dispersion-polarity choice \eqref{eq:polarity_choice} for admissible parameter sets.

\begin{figure*}[ht!]
\centering
\includegraphics[width=0.80\textwidth]{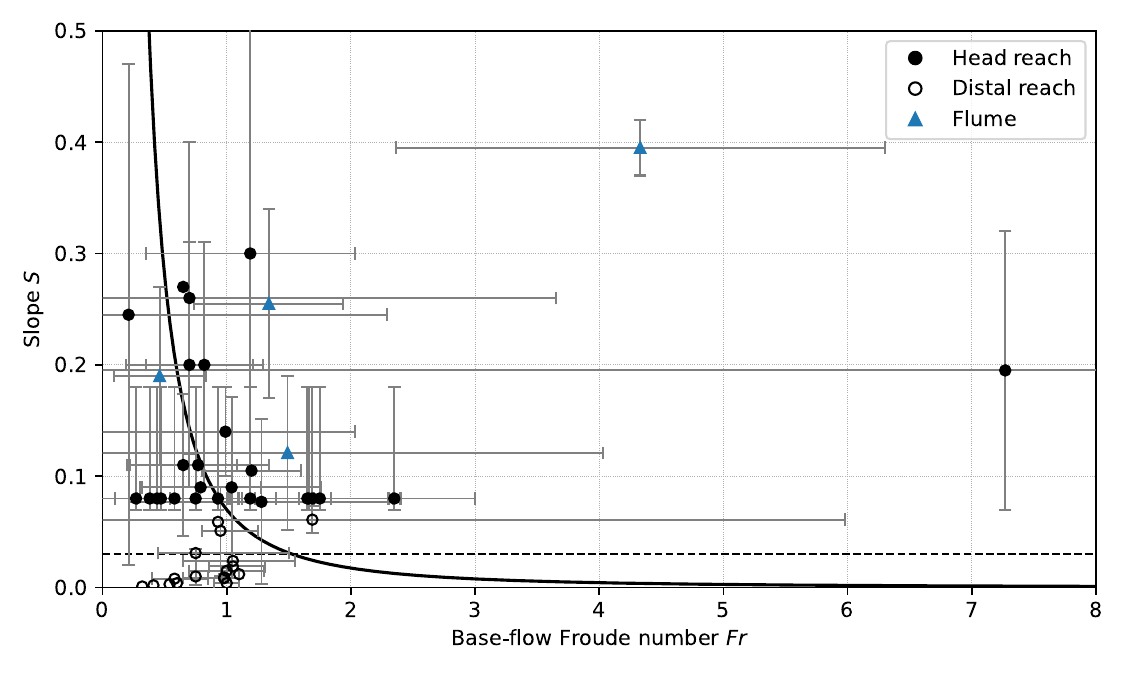}
\caption[Regime diagram in $(\mathrm{Fr},S)$ space]{%
Froude–slope $(\mathrm{Fr},S)$ regime diagram for the compiled cases. The solid curve
$S_{\mathrm{rw}}=0.07/\mathrm{Fr}^{2}$
is a guide to the eye for this dataset—an empirical organizer, not a universal threshold nor a direct consequence of linear Saint–Venant roll-wave theory. Boundaries can shift with rheology, bed roughness, channel geometry, and boundary conditions. The horizontal dashed line at $S=0.03$ marks the lower bound of the “dispersive-pulse corridor” used here to contextualize gentle reaches; it is likewise dataset-specific. Symbols: solid circles = head-reach field points; open circles = distal field points; solid triangles = laboratory flume runs (see legend). Numerical ranges and sources for all symbols are listed in Table~\ref{tab:merged_fr_s}. Here $\mathrm{Fr}$ is the event-scale Froude number computed from published $u$–$h$ pairs; $\mathrm{Fr}_0$ is reserved for the base-flow value used in the asymptotics and in the nonlinearity proxy $\tilde{\alpha}_{\mathrm{norm}}$ (Figure~\ref{fig:alphaFr}). Interpretation note: this panel provides regime context; “KdV-admissible” in the text additionally requires the dispersion-polarity choice $\beta_{\mathrm{eff}}>0$ and a positive estimated nonlinearity $\alpha_{\mathrm{eff}}^{\mathrm{lab}}>0$ (or proxy sign when $V$ is unavailable). Heterogeneous measurement methods preclude uniform error bars, so points should be viewed as representative rather than precise thresholds.}
\label{fig:regime}
\end{figure*}

\begin{table*}[ht!]
\centering
\caption{Compilation of slope $S$ and Froude number $\mathrm{Fr}$
used in this study. Reach
classes: \textit{head} ($S_{\text{mid}}>0.07$), \textit{distal}
($S_{\text{mid}}\le 0.07$), \textit{lab} = flume. For rows A1--E4, $\mathrm{Fr}$ was derived from the $u$–$h$ data listed in Table~IV of Rickenmann~\cite{Rickenmann1999} using the formula $\mathrm{Fr} = u/\sqrt{g h}$ with $g = 9.81~\text{m s}^{-2}$.}
\label{tab:merged_fr_s}
\small
\begin{tabular}{@{}lccccc@{}}
\toprule
ID & Site / Event & Reach &
$S$ (\emph{as-published}) &
$\mathrm{Fr}$ (\emph{derived or as-published}) &
Reference \\[2pt]
\midrule
A1  & T.\ Moscardo (Italy)               & Head   & $0.110\pm0.000$ & $0.77\pm0.57$ & Arattano et al 1996 \cite{Rickenmann1999} \\
A2  & Kamikamihori (Japan)               & Head   & $0.090\pm0.000$ & $0.79\pm0.49$ & Okuda \& Suwa 1981  \cite{Rickenmann1999}  \\
A3  & Mt St Helens - Shoestring (US)     & Head   & $0.260\pm0.140$ & $0.70\pm2.95$ & Pierson 1986  \cite{Rickenmann1999}   \\
B   & Jiangjia Gully (China)             & Distal & $0.061\pm0.012$ & $1.69\pm4.29$ & Jakob \& Wang (p.c.)   \cite{Rickenmann1999}   \\
C   & Swiss Alps (multiple)              & Head   & $0.300\pm0.230$ & $1.19\pm0.84$ & VAW 1992, Zimmermann 1996  \cite{Rickenmann1999}       \\
D1  & Mt St Helens - Pine C./Muddy R.    & Head   & $0.077\pm0.074$ & $1.28\pm1.07$ & Pierson 1985  \cite{Rickenmann1999}   \\
D2  & Nevado del Ruiz (Colombia)         & Head   & $0.090\pm0.081$ & $1.04\pm0.72$ & Pierson et al 1990  \cite{Rickenmann1999}   \\
E1  & Flume - Jiangjia (CN)              & Lab    & $0.121\pm0.069$ & $1.49\pm2.54$ & Wang \& Zhang 1990  \cite{Rickenmann1999}      \\
E2  & Flume - New Zealand (NZ)           & Lab    & $0.190\pm0.080$ & $0.46\pm0.37$ & Davies 1994  \cite{Rickenmann1999}    \\
E3  & Flume - Conveyor belt (NZ)         & Lab    & $0.255\pm0.085$ & $1.34\pm0.60$ & Davies 1990  \cite{Rickenmann1999}    \\
E4  & Flume - Garcia Aragon (US)         & Lab    & $0.395\pm0.025$ & $4.33\pm1.97$ & Garcia Aragon 1996  \cite{Rickenmann1999}    \\[2pt]
\midrule
R1  & Rio Reventado (CR)                 & Head   & $0.110\pm0.064$ & $0.65\pm0.43$ & Waldron 1967 \cite{Costa1984}   \\
R2  & Hunshui Gully (CN)                 & -      & (slope missing)               & $1.84\pm0.49$ & Li \& Luo 1981 \cite{Costa1984}     \\
R3  & Bullock Creek (NZ)                 & Head   & $0.105\pm0.000$ & $1.20\pm0.40$ & Pierson 1981 \cite{Costa1984}   \\
R4  & Pine Creek (WA, US)                & Head   & $0.195\pm0.125$ & $7.27\pm12.47$& Fink et al 1981 \cite{Costa1984}      \\
R5  & Mt St Helens - Morton/Campbell (US)& Head   & $0.200\pm0.110$ & $0.70\pm0.51$ & Morton \& Campbell 1974 \cite{Costa1984}    \\
R6  & Wrightwood Canyon (CA, US)         & Head   & $0.200\pm0.110$ & $0.82\pm0.47$ & Sharp \& Nobles 1953 \cite{Costa1984}     \\
R7  & Lesser Almatinka (KZ)              & Head   & $0.140\pm0.040$ & $0.99\pm1.04$ & Niyazov \& Degovets 1975 \cite{Costa1984}   \\
R8  & Matanuska Glacier (AK, US)         & Head   & $0.245\pm0.225$ & $0.21\pm2.08$ & Lawson 1982 \cite{Costa1984}    \\
R9  & Nojiri River (Japan)              & Head   &
$0.075\pm0.017$ & $2.68\pm0.06$ & Watanabe \& Ikeya 1981 \cite{Costa1984} \\ 
R10 & Mayflower Gulch (CO, US)           & Head   & $0.270\pm0.000$ & $0.65\pm0.00$ & Curry 1966 \cite{Costa1984}     \\
R11 & Dragon Creek (AZ, US)              & Distal & $0.059\pm0.000$ & $0.93\pm0.00$ & Cooley et al 1977 \cite{Costa1984}    \\[4pt]
\midrule
B1  & Illgraben 21 Jun 2019 (CH)         & Head   & 0.08 (0.07-0.18) & $1.19\pm0.65$ & Table 1 of Bolliger \cite{Bolliger2023} \\
B2  & Illgraben 02 Jul 2019 (CH)         & Head   & 0.08 (0.07-0.18) & $0.93\pm0.65$ & Table 1 of Bolliger \cite{Bolliger2023} \\
B3  & Illgraben 26 Jul 2019 (CH)         & Head   & 0.08 (0.07-0.18) & $2.35\pm0.65$ & Table 1 of Bolliger \cite{Bolliger2023} \\
B4  & Illgraben 11 Aug 2019 (CH)         & Head   & 0.08 (0.07-0.18) & $1.65\pm0.65$ & Table 1 of Bolliger \cite{Bolliger2023} \\
B5  & Illgraben 20 Aug 2019 (CH)         & Head   & 0.08 (0.07-0.18) & $0.27\pm0.65$ & Table 1 of Bolliger \cite{Bolliger2023} \\
B6  & Illgraben 24 Jun 2021 (CH)         & Head   & 0.08 (0.07-0.18) & $1.69\pm0.65$ & Table 1 of Bolliger \cite{Bolliger2023} \\
B7  & Illgraben 06 Jul 2021 (CH)         & Head   & 0.08 (0.07-0.18) & $1.75\pm0.65$ & Table 1 of Bolliger \cite{Bolliger2023} \\
B8  & Illgraben 16 Jul 2021 (CH)         & Head   & 0.08 (0.07-0.18) & $0.58\pm0.65$ & Table 1 of Bolliger \cite{Bolliger2023} \\
B9  & Illgraben 07 Aug 2021 (CH)         & Head   & 0.08 (0.07-0.18) & $0.47\pm0.65$ & Table 1 of Bolliger \cite{Bolliger2023} \\
B10 & Illgraben 19 Sep  2021 (CH)         & Head   & 0.08 (0.07-0.18) & $0.38\pm0.65$ & Table 1 of Bolliger \cite{Bolliger2023} \\
B11 & Illgraben 05 Jun 2022 (CH)         & Head   & 0.08 (0.07-0.18) & $0.75\pm0.65$ & Table 1 of Bolliger \cite{Bolliger2023} \\
B12 & Illgraben 04 Jul 2022 (CH)         & Head   & 0.08 (0.07-0.18) & $1.66\pm0.65$ & Table 1 of Bolliger \cite{Bolliger2023} \\
B13 & Illgraben 08 Sep 2022 (CH)         & Head   & 0.08 (0.07-0.18) & $0.44\pm0.65$ & Table 1 of Bolliger \cite{Bolliger2023} \\[2pt]
\midrule
G1  & OR Beach (lab rill, US)            & Distal & 0.015 (0.012-0.018) & 1.0 (0.7-1.3) & Fig 4 of Grant \cite{Grant1997} \\
G2  & Pasig-Potrero (PH)                 & Distal & 0.012                 & 1.1           & Fig 4 of Grant \cite{Grant1997} \\
G3  & Govers rill (unpubl.)              & Distal & 0.051 (0.01-0.09)     & 0.95 (0.8-1.25) & Fig 4 of Grant \cite{Grant1997} \\
G4  & Inbar \& Schick 1979 (IL)          & Distal & 0.031 (0.029-0.035)   & 0.75 (0.45-1.5) & Fig 4 of Grant \cite{Grant1997} \\
G5  & Fahnestock 1963 (US)               & Distal & 0.024                 & 1.05 (0.65-1.55) & Fig 4 of Grant \cite{Grant1997} \\
G6  & Grimm \& Leupold 1939 (DE)         & Distal & 0.019                 & 1.05 (0.86-1.31) & Fig 4 of Grant \cite{Grant1997} \\
G7  & Milhous 1973 (US)                  & Distal & 0.010                 & 0.75           & Fig 4 of Grant \cite{Grant1997} \\
G8  & Laronne et al.\ 1986 (IL)          & Distal & 0.009                 & 0.98           & Fig 4 of Grant \cite{Grant1997} \\
G9  & Pitlick 1992 (US)                  & Distal & 0.0085 (0.007-0.012)  & 0.98 (0.65-1.1) & Fig 4 of Grant \cite{Grant1997} \\
G10 & Ritter 1967 (US)                   & Distal & 0.008                 & 0.58 (0.4-0.7) & Fig 4 of Grant \cite{Grant1997} \\
G11 & Dinehart 1992 (US)                 & Distal & 0.004                 & 0.60           & Fig 4 of Grant \cite{Grant1997} \\
G12 & Pierson \& Scott 1985 (US)         & Distal & 0.004                 & 1.0 (0.9-1.1) & Fig 4 of Grant \cite{Grant1997} \\
G13 & Butler 1977 (US)                   & Distal & 0.003                 & 0.54           & Fig 4 of Grant \cite{Grant1997} \\
G14 & Emmett 1976 (US)                   & Distal & 0.002                 & 0.41           & Fig 4 of Grant \cite{Grant1997} \\
G15 & Burrows et al.\ 1979 (US)          & Distal & 0.001                 & 0.32           & Fig 4 of Grant \cite{Grant1997} \\
\bottomrule
\end{tabular}
\end{table*}

\subsection{Assumptions and Validity of the KdV Model}

Formulating a KdV equation for debris flows necessitates several simplifying assumptions that delimit the parameter space in which coherent wave structures (e.g., cnoidal waves or solitons) can form. It is important to emphasize that the present formulation is most applicable to water-rich, fluidized debris flows where surge front waves emerge as localized, high-energy pulses. In this context, the soliton or cnoidal wave represents a localized wavefront that enhances mass transport rather than explaining long runout solely based on large volume or dry conditions.

We make the standard weakly nonlinear, long-wave assumptions: small relative free-surface amplitude $a/h_0=\epsilon\ll1$ and a large wavelength-to-depth ratio $L_{\mathrm{w}}/h_0\gg1$. These constraints ensure that the leading balance between nonlinear steepening and curvature-induced dispersion is captured by a KdV-type reduction of the depth-averaged equations.

One of the key approximations used in our calculation is that the internal (normal) stress varies linearly with the free-surface curvature, i.e., $\tilde{\tau}_{xx} = \gamma \eta_{xx}$, where $\tilde{\tau}_{xx}$ is a dimensionless measure of stress, $\eta(x,t)$ is the free-surface perturbation, and $\gamma$ scales an effective elastic or dispersive response. This linearization holds for (i) $a \ll h_0$, i.e., the wave amplitude is small relative to the mean flow depth $h_0$, and (ii) $L_{\mathrm{w}} \gg h_0$, i.e., the characteristic wavelength $L_{\mathrm{w}}$ is much larger than $h_0$. Under these conditions, higher-order stress–strain effects can be omitted, and the flow remains shallow over long distances (e.g., runout zones with low slopes or near horizontal or fluidized debris flows). Consequently, the dominant dispersive term in the momentum equation is of the type $\beta \partial_x^3 \eta$, and steeper gradients or shorter wavelengths are omitted. This ensures that the leading-order balance between nonlinear steepening and dispersion is tractable within the KdV model.

The quadratic coefficient $\alpha$ is not fixed by a single constitutive closure. When crest-celerity data are available, we estimate an effective laboratory-frame coefficient from the amplitude–celerity relation $V=u_0+c_0+(\alpha_{\mathrm{eff}}^{\mathrm{lab}}A)/3$ (with $A=h_{1s}-h_0$ and $c_0$ as defined in §~\ref{sec:notation}), and compare it to the shallow-water baseline $\alpha_0^{\mathrm{lab}}=3c_0/(2h_0)$ via $\alpha_{\mathrm{norm}}=\alpha_{\mathrm{eff}}^{\mathrm{lab}}/\alpha_0^{\mathrm{lab}}$.
When only depth-averaged surge velocities are available, we also report a velocity-based proxy $\tilde{\alpha}_{\mathrm{eff}}^{\mathrm{lab}}=3(u_{1s}-u_0-c_0)/A$ and its normalized form $\tilde{\alpha}_{\mathrm{norm}}= \tilde{\alpha}_{\mathrm{eff}}^{\mathrm{lab}}/\alpha_0^{\mathrm{lab}}$ for descriptive plotting; because $u_{1s}$ is not the crest speed, this proxy can under-estimate $\lvert\alpha_{\mathrm{eff}}^{\mathrm{lab}}\rvert$ and may misclassify signs near zero, so celerity-based values are preferred when $V$ is known.
With the dispersion polarity used here ($\beta_{\mathrm{eff}}>0$), elevation-soliton admissibility corresponds to $\alpha_{\mathrm{eff}}^{\mathrm{lab}}>0$; negative values are consistent with depression-like/roll-wave behavior. This estimation strategy avoids ad hoc slow-time scalings and is consistent with the weakly nonlinear ordering adopted in the derivation.

We adopt a thresholded basal resistance consistent with a depth-averaged Bingham closure: once the applied shear exceeds the yield stress $\tau_y$, the basal stress is modeled as $\tau_b=\tau_y+\mu_{\mathrm{eff}}u$ with $\mu_{\mathrm{eff}}=\eta_{\mathrm{3D}}/h$. In the weakly nonlinear, long-wave derivation this mobilization renders the leading KdV balance independent of $\tau_y$; $\tau_y$ influences onset (and weak damping) rather than the $\mathcal{O}(\epsilon)$ coefficients. Many debris-flow materials are better represented by Herschel–Bulkley or related laws; such refinements renormalize the effective coefficients $(\alpha,\beta)$ but leave the basic balance between weak nonlinearity and dispersion, and thus the admissibility of KdV-type pulses, intact.

Granular friction can depend on depth and shear rate, $\mu=\mu(h,u,\dot{\gamma},\ldots)$; see \cite{pouliquen2002friction}. In this setting the basal shear may be written
\begin{equation}
\tau_b=\mu(h,\mathrm{Fr}) \rho g h \cos\theta \operatorname{sgn}(u), \quad \mathrm{Fr}=\frac{|u|}{\sqrt{g h}}.
\end{equation}
The base balance changes from $S=\mu_s$ to
\begin{equation}
S=\mu(h_0,\mathrm{Fr}_0)
\end{equation}
where $S\equiv\tan\theta$ by definition; on gentle slopes $\sin\theta\simeq S$ and $\cos\theta\simeq 1$. Small perturbations introduce sensitivities $\mu_h=\partial\mu/\partial h$ and $\mu_{\mathrm{Fr}}=\partial\mu/\partial \mathrm{Fr}$ that vary slowly on the wave scale. Under the weakly nonlinear, long-wave ordering ($k h_0\ll 1$), the KdV core persists with dispersion fixed by the curvature closure, while the linear speed $c$ and the quadratic nonlinearity $\alpha$ become slowly varying. A weak KdV–Burgers correction appears when the base is not exactly balanced:
\begin{multline}
\eta_t + c(x,t) \eta_x + \alpha(x,t) \eta \eta_x + \beta_{\mathrm{eff}} \eta_{xxx}
= \partial_x \big[D(x,t) \eta_x\big] \\
+ \text{higher-order terms}.
\end{multline}
Here $\beta_{\mathrm{eff}}$ is unaffected by the friction law (it follows from the curvature closure), whereas $c$ and $\alpha$ inherit slow variability through $\mu(h,\mathrm{Fr})$, and $D$ summarizes the near-balance mismatch. Provided $\mu$ varies slowly on the wave scale and $\mathrm{Fr}_0$ lies within the empirical calibration range, solitary and cnoidal pulses remain admissible with the same dispersion polarity ($\beta_{\mathrm{eff}}>0$).

Finally, assuming $\epsilon \ll 1$ allows us to truncate higher-order ($\mathcal{O}(\epsilon^2)$ and beyond) terms. This isolates the leading-order balance between nonlinearity and dispersion, yielding a KdV-type equation. While our formulation suits fluidized, water-rich debris flows with surge-front dynamics, large dry landslides or flows on steep slopes may invoke strong shocks, additional yield stresses, or other processes not captured by this approach. Hence, the present KdV-based model should be viewed as a heuristic for surge fronts in relatively fluidized debris flows, rather than a comprehensive solution for all long-runout landslide phenomena.

In Appendix~\ref{app:coulomb}, we show that when basal resistance follows a Coulomb law the weakly nonlinear, long-wave multiple-scale expansion produces the same leading-order KdV balance: $\alpha$ and $\beta$ at $\mathcal{O}(\epsilon)$ are unchanged and Coulomb friction enters only as a higher-order weak damping. Within the ranges considered, this leaves the admissibility criterion and regime map unchanged at leading order.

To keep the observational context close to the modeling assumptions, we next juxtapose the regime analysis with the compiled field and laboratory measurements. The $(\mathrm{Fr},S)$ diagram and Table~\ref{tab:merged_fr_s} are therefore discussed immediately in the following subsection. Here the event-scale Froude number is defined as
\begin{equation}
\mathrm{Fr} = \frac{u}{\sqrt{g h}},
\end{equation}
where $u$ is the depth-averaged velocity ($\text{m s}^{-1}$), $h$ is the flow thickness (m), and $g$ is gravitational acceleration ($9.81~\text{m s}^{-2}$). These $(u,h)$ pairs are extracted from published field or laboratory measurements. We reserve $\mathrm{Fr}_0$ for the uniform base-flow Froude number used in the asymptotic derivation and in defining $\alpha_{\mathrm{norm}}$.

\subsection{Roll--wave versus dispersive--pulse regimes}
\label{sec:roll-vs-disp}

Roll waves, first analyzed in the context of shallow open-channel flow by Dressler~\cite{Dressler1949}, arise when inertial steepening is balanced primarily by basal drag without the need to invoke surface-curvature dispersion. In debris-flow settings, depth-averaged momentum-integral models reproduce these strongly nonlinear, shock-like surge fronts on sufficiently steep reaches, and connect their stability to the local Froude number and slope~\cite{Balmforth2004}.

Field monitoring at instrumented torrents shows that individual events often comprise multiple surges with structured fronts; high-cadence 3-D LiDAR and other sensors at Illgraben resolve these trains clearly in downstream reaches where the bed slope decreases relative to the head channel~\cite{Schimmel2022,Spielmann2024, aaron2023high,zhang2021analyzing, walter2023seismic}. Similarly, velocity-profiling measurements in the Moscardo system document depth-resolved surge kinematics, but do not by themselves establish a precise low-slope threshold for pulse amplitudes~\cite{arattano2012analysis}. To avoid over-interpretation, we therefore refer to these downstream segments as ``gentler'' or ``moderate-slope'' reaches rather than assigning a single numerical cutoff.

On such gentler segments, curvature-related normal stresses can enter the leading-order balance alongside nonlinear advection, so that a KdV-type description with weakly dispersive pulses becomes appropriate; on steeper head reaches, classic roll-wave behavior remains the correct description. In this sense, roll waves and dispersive pulses are complementary, regime-selective mechanisms within the same event, with the active balance set by local $\mathrm{Fr}$ and bed slope (quantified in Figure~\ref{fig:regime}).

The regime map introduced below aggregates published field points and laboratory runs into the $(\mathrm{Fr},S)$ plane to illustrate this partition: a shock-dominated domain consistent with roll-wave theory at higher slopes and a corridor on gentler terrain where dispersive pulses are admissible under the depth-averaged KdV reduction. We emphasize that the corridor bounds reflect consistency across data sets and theory rather than a universal site-independent threshold; site geometry, material properties, and pore-fluid effects can shift these bounds within the same regime map.

Accordingly, in Figure~\ref{fig:regime} we plot an empirical guide $S_{\mathrm{rw}}(\mathrm{Fr}) \equiv 0.07/\mathrm{Fr}^{2}$, where the subscript “rw” denotes “roll-wave.” The coefficient $0.07$ was chosen to pass through the steep-slope/roll-wave cluster in our compilation and serves only as an organizer for this dataset; it is not a universal threshold and not a direct result of linear Saint--Venant roll-wave theory (cf.\ Dressler~\cite{Dressler1949}). Classical linear stability usually states onset in terms of a critical Froude number (often $\mathcal{O}(2)$) rather than the specific inverse-quadratic $S_{\mathrm{rw}}(\mathrm{Fr})$ form, and site-specific friction and geometry can shift any apparent boundary.

\subsection{Nonlinearity versus Base-Flow Froude Number: Two Regimes in Debris-Flow Surge Dynamics\label{sec:DebrisFlowAnalysis}}

Here, we examine how an effective quadratic nonlinearity scales with base flow.  In the KdV model (Eq.~\eqref{eq:kdv_final_lab}), the laboratory-frame speed–amplitude relation is
\begin{equation}
V \approx u_0 + c_0 + \frac{\aefflab A}{3},
\quad
A=h_{1s}-h_0.
\end{equation}
Here $A$ is the elevation amplitude and $c_0$ is as defined in §~\ref{sec:notation}.
Estimating $\aefflab$ ideally requires the crest celerity $V$. Many records lack a reliable $V$. To place all events on a common axis, a velocity-based proxy is used in which $V$ is replaced by the first-surge depth-averaged velocity $u_{1s}$ (with $A=h_{1s}-h_0$):
\begin{equation}
\aefflabtilde=\frac{3 (u_{1s}-u_0-c_0)}{A},
\quad
\alpha_0^{\mathrm{lab}}=\frac{3c_0}{2h_0},
\quad
\anormtilde=\frac{\aefflabtilde}{\alpha_0^{\mathrm{lab}}}.
\end{equation}
Because $u_{1s}$ is a depth-averaged speed rather than the crest speed, one typically has $u_{1s}\lesssim V$ for elevation surges; the proxy therefore tends to underestimate $\aefflab$, and small negative values can arise from this substitution and measurement noise. The proxy is descriptive only; the KdV derivation, dispersion-sign choice, distal-work budget, and numerical solutions do not depend on it. Where $V$ is available, one may compute
\begin{equation}
\aefflab=\frac{3 (V-u_0-c_0)}{A},
\quad
\anorm=\frac{\aefflab}{\alpha_0^{\mathrm{lab}}},
\end{equation}
and compare with the proxy-based values.

We note that where crest celerity $V$ is unavailable we use the velocity-based proxy $\aefflabtilde$ and its normalized form $\anormtilde$; because $u_{1s}\lesssim V$, this proxy tends to underestimate $\aefflab$ and can flip sign near zero. For admissibility and trend fits we treat $|\anorm|<10^{-3}$ or $|\anormtilde|<10^{-3}$ as near-zero/indeterminate and report such entries in scientific notation (e.g., $1\times10^{-4}$) rather than as $0.00$.

\begin{figure}[ht!]
\centering
\includegraphics[width=\linewidth]{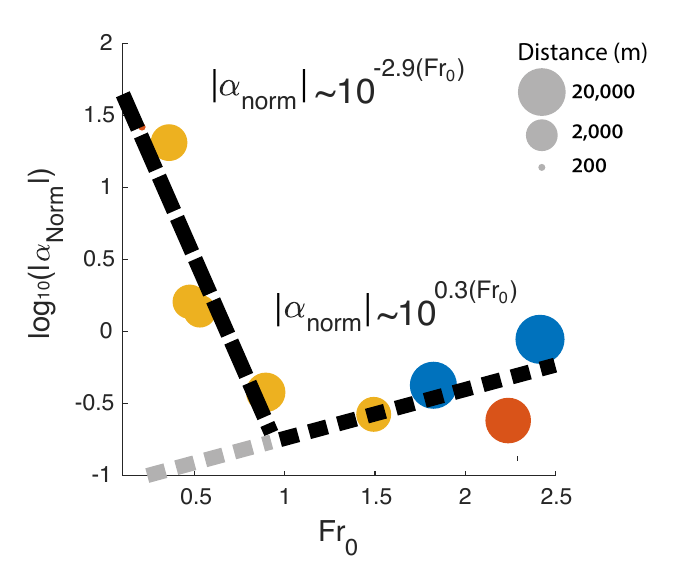} 
\caption{ Semilog plot of $\log_{10} |\tilde{\alpha}_{\mathrm{norm}}|$ versus base-flow Froude number $\mathrm{Fr}_0=u_0/\sqrt{g h_0}$. The plotted quantity is a velocity-based proxy with $\tilde{\alpha}_{\mathrm{eff}}^{\mathrm{lab}}=3(u_{1s}-u_0-c_0)/A$, $\alpha_0^{\mathrm{lab}}=3c_0/(2h_0)$, and $\tilde{\alpha}_{\mathrm{norm}}=\tilde{\alpha}_{\mathrm{eff}}^{\mathrm{lab}}/\alpha_0^{\mathrm{lab}}$, where $A=h_{1s}-h_0$ and $c_0$ as defined in §~\ref{sec:notation}. Markers show $|\tilde{\alpha}_{\mathrm{norm}}|$; filled markers indicate $\tilde{\alpha}_{\mathrm{eff}}^{\mathrm{lab}}<0$. Straight-line guides summarize apparent trends: for $\mathrm{Fr}_0<1$, $\log_{10}|\tilde{\alpha}_{\mathrm{norm}}| \approx -2.9 \mathrm{Fr}_0$; for $\mathrm{Fr}_0>1$, $\log_{10}|\tilde{\alpha}_{\mathrm{norm}}| \approx 0.3 \mathrm{Fr}_0$. These lines are derived from semi-log linear regressions between $\log_{10} |\tilde{\alpha}_{\mathrm{norm}}|$ and $\mathrm{Fr}_0$. Negative proxy values commonly occur when $u_{1s}<u_0+c_0$ for $A>0$; because the proxy substitutes $u_{1s}$ for the crest speed $V$, it can underestimate $\alpha_{\mathrm{eff}}^{\mathrm{lab}}$, so proxy sign alone is not used as a KdV-admissibility criterion (celerity-based estimates are preferred where available). One sample from Rio Moscardo (20~Jul~1993) is not included in the analysis due to a very small magnitude close to zero (e.g., $<10^{-3}$).}
\label{fig:alphaFr}
\end{figure}

Figure~\ref{fig:alphaFr} plots $\log_{10} \bigl|\tilde{\alpha}_{\mathrm{norm}}\bigr|$ versus the base-flow Froude number $\mathrm{Fr}_0=u_0/\sqrt{g h_0}$ (semilog: linear $x$, logarithmic $y$) based on the available data for debris flows with multiple surge events (Table \ref{tab:debris_flow_table}, recalculated using compiled data from~\cite{zanuttigh2007instability}). Marker fill encodes the sign of $\tilde{\alpha}_{\mathrm{eff}}^{\mathrm{lab}}$ (filled for $<0$, open for $>0$); the ordinate shows the magnitude via $|\cdot|$, so sign information is conveyed only by the marker fill. Two straight lines derived from semi-log linear regression summarize the apparent trends:
\begin{align}
\bigl|\tilde{\alpha}_{\mathrm{norm}}\bigr| \sim & 10^{-2.9 \mathrm{Fr}_0}\quad(\mathrm{Fr}_0<1),  \nonumber \\
\bigl|\tilde{\alpha}_{\mathrm{norm}}\bigr| \sim & 10^{0.3 \mathrm{Fr}_0} \quad(\mathrm{Fr}_0>1).
\end{align}
In the subcritical regime ($\mathrm{Fr}_0<1$) the proxy decreases strongly with $\mathrm{Fr}_0$; in the supercritical regime ($\mathrm{Fr}_0>1$) it varies weakly. Filled markers (negative proxy) typically occur when $u_{1s}<u_0+c_0$ for $A>0$, consistent with depression-like behavior or roll-wave dominance, but small negative proxy values can also arise because the proxy substitutes the depth-averaged speed $u_{1s}$ for the crest speed $V$ and thus tends to underestimate $\alpha_{\mathrm{eff}}^{\mathrm{lab}}$. The lines are descriptive only, not universal scalings. KdV-admissibility is judged from the broader context (slope/Froude corridor, amplitude range, dispersion sign) and, where available, the celerity-based $\anorm$ rather than from the proxy sign alone. Very small magnitudes are reported in scientific notation (e.g., $10^{-4}$) rather than as $0.00$ in the Table~\ref{tab:debris_flow_table}. One sample from Rio Moscardo (20~Jul~1993) was excluded from the analysis above because its magnitude is indistinguishable from zero within uncertainty.

The diagnostic is evaluated in the laboratory frame. Within the inviscid KdV core, the coefficients $\alpha$ and $\beta$ are invariant under Galilean shifts; weak dissipation is treated as a small KdV–Burgers correction and handled in the full-order numerics (Sec.~\ref{sec:weak_damping}). For field-based soliton tagging we use a pragmatic weak-nonlinearity window $0.05\le \epsilon=(h_{1s}-h_0)/h_0 \le 0.30$ and require elevation polarity, $\aefflab>0$ (equivalently $\anorm>0$); the dispersion polarity $\beta_{\mathrm{eff}}>0$ is enforced via the curvature-closure choice $\gamma>0$ (Sec.~\ref{sec:dispersion}). In the numerical tests, to minimize higher-order contamination we use smaller initial amplitudes, $0.05\le \epsilon \le 0.15$ (see Methodology, “Initial conditions”). Values with $|\anorm|<10^{-3}$ are treated as near-zero/indeterminate and excluded from admissibility and trend fits; in the tables we report such cases in scientific notation (e.g., $1\times10^{-4}$) rather than as $0.00$.

\begin{table*}[ht!]
\centering
\caption{Available data for debris flows with multiple surge events from \protect\cite{zanuttigh2007instability}. Columns include event duration (TE), base depth $h_0$, base velocity $u_0$, first-surge depth $h_{1s}$, first-surge depth-averaged velocity $u_{1s}$, advection distance (Dist.), base-flow Froude $\mathrm{Fr}_0=u_0/\sqrt{g h_0}$, and the dimensionless amplitude $\epsilon=(h_{1s}-h_0)/h_0$. The nonlinearity columns use a velocity-based proxy: the entry labeled $\alpha_{1s}$ corresponds to $\tilde{\alpha}_{\mathrm{eff}}^{\mathrm{lab}}=3(u_{1s}-u_0-c_0)/A$, with $\alpha_0^{\mathrm{lab}}=3c_0/(2h_0)$ and $\tilde{\alpha}_{\mathrm{norm}}=\tilde{\alpha}_{\mathrm{eff}}^{\mathrm{lab}}/\alpha_0^{\mathrm{lab}}$, where $A=h_{1s}-h_0$ and $c_0$ is as defined in §~\ref{sec:notation}. All quantities are evaluated in the laboratory frame.
\newline\newline
\footnotesize
\textit{Notes:} TE = event duration; $h_0$, $u_0$ = base-flow depth and velocity; $h_{1s}$, $u_{1s}$ = first-surge depth and depth-averaged velocity. Dist. $\approx u_0 \text{TE}$ (advection distance at the base speed; approximate).
\emph{Nonlinearity diagnostic (velocity-based proxy):}
The column $\alpha_{1s}$ lists $\tilde{\alpha}_{\mathrm{eff}}^{\mathrm{lab}}=3 (u_{1s}-u_0-c_0)/(h_{1s}-h_0)$, with $A=h_{1s}-h_0$ (taken $>0$ for elevation surges) and $c_0$ is as defined in §~\ref{sec:notation}. The shallow-water reference is $\alpha_0^{\mathrm{lab}}\equiv\alpha_0=3c_0/(2h_0)$, and the column $\tilde{\alpha}_{\mathrm{norm}}$ reports the normalized proxy $\tilde{\alpha}_{\mathrm{eff}}^{\mathrm{lab}}/\alpha_0^{\mathrm{lab}}$.
Where $A\le 0$, the proxy is not reported. Very small magnitudes are shown in scientific notation (e.g., $10^{-3}$) rather than as $0.00$. Because the proxy substitutes $u_{1s}$ for the crest speed $V$, it can underestimate $\alpha_{\mathrm{eff}}^{\mathrm{lab}}$; celerity-based estimates are preferred when $V$ is available.}
\label{tab:debris_flow_table}
\scriptsize
\begin{tabular}{@{}l l r r r r r r r r r r r p{3.2cm}@{}} 
\toprule
Field Site & Date & TE (s) & $h_0$ (m) & $u_0$ (m/s) & $h_{1s}$ (m) & $u_{1s}$ (m/s) & Dist. (m) & $\mathrm{Fr}_0$ & $\epsilon$ & $\alpha_{1s}$ & $\alpha_0$ & $\tilde{\alpha}_{\mathrm{norm}}$ & Comments \\
\midrule
Illgraben, Switzerland & 28-Jun-01 & 3600 & 0.58 & 5.763 & 1.6 & 6.3 & 20747 & 2.42 & 1.76 & -5.44 & 6.17 & -0.881 & muddy, $>$30 surges \\
Illgraben, Switzerland & 28-Jun-00 & 3600 & 0.51 & 4.08 & 2.9 & 4.1 & 14688 & 1.82 & 4.69 & -2.78 & 6.58 & -0.423 & granular, 17 surges \\
Acquabona, Italy & 17-Aug-98 & 2280 & 0.52 & 5.058 & 1.7 & 6.7 & 11532 & 2.24 & 2.27 & -1.57 & 6.52 & -0.241 & 20 surges \\
Acquabona, Italy & 27-Jul-98 &  420 & 1.40 & 0.765 & 1.5 & 1.0 &   321 & 0.21 & 0.07 & -104.13 & 3.97 & -26.224 & single surge \\
Rio Moscardo, Italy & 08-Jul-96 &  900 & 1.68 & 2.142 & 3.0 & 4.0 &  1928 & 0.53 & 0.79 & -5.00 & 3.62 & -1.380 & roll waves \\
Rio Moscardo, Italy & 22-Jun-96 &  900 & 0.36 & 2.805 & 2.5 & 3.2 &  2525 & 1.49 & 5.94 & -2.08 & 7.83 & -0.266 & 3 surges \\
Rio Moscardo, Italy & 14-Sep-93 & 2940 & 0.95 & 1.088 & 1.0 & 2.5 &  3199 & 0.36 & 0.05 & -98.45 & 4.82 & -20.424 & roll waves \\
Rio Moscardo, Italy & 20-Jul-93 & 1680 & 0.35 & 2.448 & 2.3 & 4.3 &  4113 & 1.32 & 5.57 & $-1.5\times10^{-3}$ & 7.94 & $-1.9\times10^{-4}$ & 3 surges and roll waves \\
Rio Moscardo, Italy & 11-Jul-93 & 2280 & 0.51 & 1.998 & 2.0 & 3.0 &  4555 & 0.89 & 2.92 & -2.49 & 6.58 & -0.378 & 2 surges and roll waves \\
Rio Moscardo, Italy & 30-Sep-91 & 1560 & 1.05 & 1.513 & 2.2 & 1.9 &  2360 & 0.47 & 1.10 & -7.36 & 4.58 & -1.606 & roll waves \\
\bottomrule
\end{tabular}
\end{table*}

\section{Numerical Validation of Cnoidal Waves and Solitons}
\label{sec:fidelity}

We test the KdV reduction by solving the full depth-averaged rheological system in the weak-amplitude, long-wave regime and asking whether it supports the same families of traveling waves (cnoidal trains and solitary pulses) with the speeds and widths predicted by the KdV model. The target flows are water-rich, fluidized debris surges for which the amplitude ratio $\epsilon=a/h_0$ is small and the dominant wavelength greatly exceeds $h_0$.

Numerical scheme, boundary conditions, initialization, and coefficient choices are detailed in the following Methodology subsection (§~\ref{sec:methodology}); throughout, we enforce the dispersion-polarity choice $\gamma>0$ (Eq.~\eqref{eq:polarity_choice}) and monitor discrete mass and depth-averaged momentum budgets, which remain closed to truncation error. We compare predicted and measured pulse speeds and quantify agreement using phase-speed error, amplitude error, and a shape-misfit metric; grid-convergence checks confirm the expected orders.

To remain within the domain of validity of the asymptotics, all runs satisfy $\epsilon\ll 1$, base-flow Froude numbers $\mathrm{Fr}_{0}$ and slopes consistent with the regime map, and $(\alpha,\beta)$ computed from the same constitutive parameters used in the solver. Where field-derived slopes are moderately steep, we restrict attention to reaches and parameter sets where the dispersive–nonlinear balance is admissible; outside that set, roll-wave models are more appropriate and the KdV reduction is not invoked. Within its envelope of assumptions, the full equations reproduce KdV-type cnoidal trains and solitary pulses with small phase and shape errors, supporting the use of the KdV model as a predictive description of surge-front propagation.

\subsection{Methodology}
\label{sec:methodology}

We solve the full depth-averaged debris-flow system in one horizontal dimension with a conservative finite-volume scheme (second-order MUSCL reconstruction) and a strong-stability-preserving third-order Runge–Kutta time integrator. The dispersive normal-stress contribution is discretized as a centered, fourth-order finite-difference operator for the third spatial derivative, and a very weak Kreiss–Oliger filter damps grid-scale noise without altering resolved phase speeds. Basal resistance follows the same closure used in the asymptotic development (Bingham: $\tau_b=\tau_y+\mu_{\mathrm{eff}} u$ with a $C^1$ regularization near $u = 0$); sensitivity runs with a Coulomb option $\tau_b=\mu_s \rho g h$ confirm that leading-order wave properties are unchanged within the parameter ranges analyzed. At every step we enforce the dispersion-polarity choice $\gamma>0$ (see Eq.~\eqref{eq:polarity_choice}).

Boundary conditions are matched to the waveform under test. For cnoidal waves we use a periodic domain to preserve spectral purity; for solitary pulses we apply radiation (Sommerfeld/characteristic) outflow with a short sponge layer to prevent spurious reflections.

Unless otherwise noted, the computational domain has length $L_{\mathrm{dom}}=100\text{ m}$ with $N_x=1000$ cells ($\Delta x=0.1\text{ m}$) and total simulated time $T=100\text{ s}$ advanced with a time step $\Delta t=0.01\text{ s}$. Here $\Delta x$ and $\Delta t$ denote the numerical grid spacing and the time-step size, respectively. The Courant number remains in the range 0.3--0.5 to ensure stability. Grid-refinement studies verify second-order convergence of the hyperbolic core and fourth-order convergence of the dispersive operator in isolation (halving $\Delta x$ and $\Delta t$).

For solitary pulses we initialize a KdV-consistent profile
\begin{equation}
\eta(x,0)=A \sech^{2} \Bigl(\frac{x-x_{0}}{L_s}\Bigr),
\quad
L_s=\sqrt{\frac{12 \beta}{|\alpha| A}} .
\end{equation}
We use small amplitudes $A=\epsilon h_0$ with $\epsilon \in[0.05,0.15]$ to remain in the weakly nonlinear regime ($\epsilon\ll1$). The predicted lab-frame phase speed is $V_{\mathrm{lab}} = u_0 + c_0 + \alpha A/3$ with $c_0$ defined in §~\ref{sec:notation}, and the polarity is chosen so that $\alpha A/\beta>0$ for the faster-than-$u_0$ branch.
(\emph{Note.} For field/regime classification we allow a broader window $0.05\le\epsilon\le0.30$ when tagging admissible cases in the compilation; the lower bound avoids noise-level amplitudes in measurements, while the upper bound marks where weak-nonlinearity ordering starts to degrade. The $0.05$–$0.15$ band here is a conservative subset used only for numerical initialization to minimize higher-order contamination.)

For cnoidal waves, we initialize the exact KdV cnoid constructed from the same $(\alpha,\beta)$ used by the solver, with wavelength $\Lambda$ chosen so that an integer number of periods fits the domain; periodic boundaries then maintain the waveform with minimal numerical dispersion.

The effective coefficients $\alpha$ and $\beta$ are computed from the same depth-averaged rheology and curvature-stress closure used in the derivation, and the dispersion-polarity choice is imposed throughout (see Eq.~\eqref{eq:polarity_choice}). To remain consistent with the regime map, simulations target parameter sets corresponding to fluidized, water-rich surges on low-to-moderate slopes where a dispersive balance is admissible; on steeper reaches where roll-wave dynamics dominate, the KdV reduction is not invoked.

We quantify fidelity via (i) phase-speed error $E_c = \lvert c_{\mathrm{num}} - c_{\mathrm{KdV}} \rvert / \lvert c_{\mathrm{KdV}} \rvert$, (ii) amplitude drift $E_A = \lvert A_{\mathrm{num}} - A \rvert / A$, and (iii) shape misfit
\begin{equation}
E_{\mathrm{sh}} = \min_{\delta x} 
\frac{\left\|\eta_{\mathrm{num}}(\cdot,t) - \eta_{\mathrm{KdV}}(\cdot - \delta x,t)\right\|_{2}}
{\left\|\eta_{\mathrm{KdV}}(\cdot,t)\right\|_{2}}.
\end{equation}
Here the subscript $\mathrm{num}$ denotes quantities measured from the full rheological numerical solution (e.g., crest-tracked speed $c_{\mathrm{num}}$, measured amplitude $A_{\mathrm{num}}$, and free-surface field $\eta_{\mathrm{num}}(x,t)$). The subscript $\mathrm{KdV}$ denotes the matched ideal KdV reference (cnoid or soliton) constructed with the same $(h_0,u_0,\alpha,\beta)$ and target amplitude $A$; its phase speed $c_{\mathrm{KdV}}$ follows the KdV model (e.g., $u_0 + c_0$ for the linear carrier, or $u_0 + c_0 + \alpha A/3$ for a soliton), and $\eta_{\mathrm{KdV}}(x,t)$ is the corresponding analytic waveform. The $L^2$ norm $\|\cdot\|_2$ is over the computational domain, and $\delta x$ is a continuous phase shift used to align the profiles (not the numerical grid spacing $\Delta x$ defined in the methodology). In addition, we monitor discrete mass and depth-averaged momentum budgets from the governing equations; departures remain at truncation-error levels in all reported runs. These diagnostics are accumulated over many wave periods (cnoids) or over 10--20 pulse widths (solitons) after initial spin-up to ensure asymptotic propagation.

\subsection{Results}

\begin{figure}[ht!]
\centering
\includegraphics[width=0.45\textwidth]{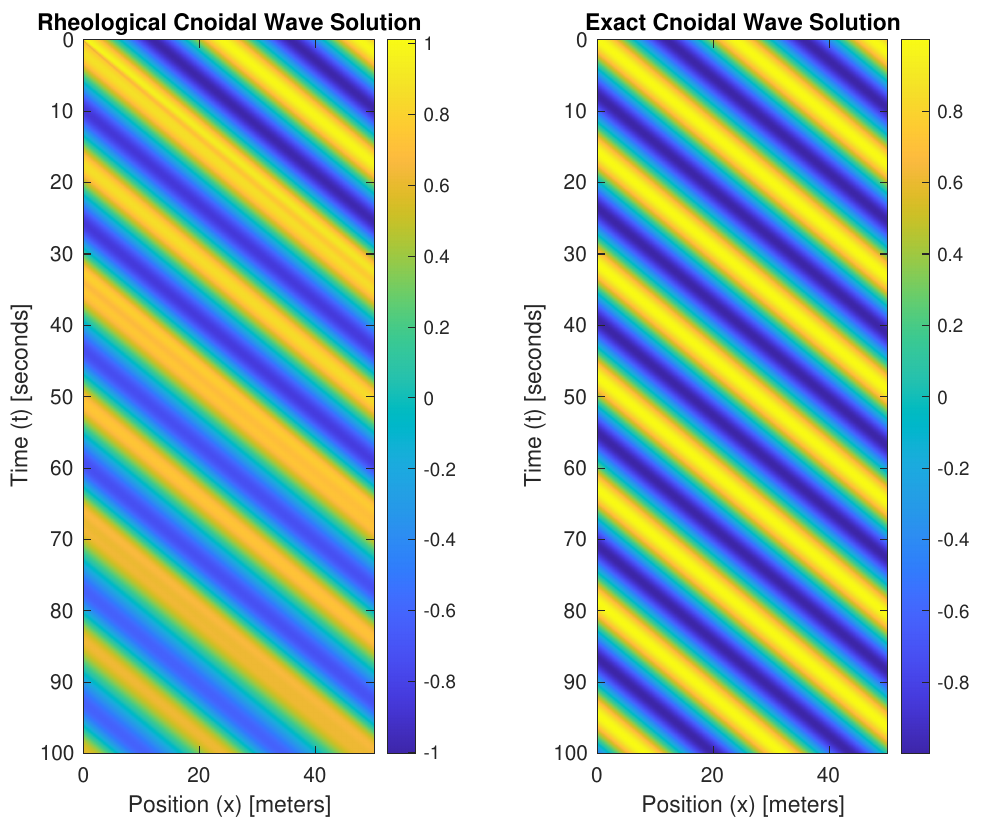}
\caption{Cnoidal wave propagation under the full rheological model (left) compared to the ideal cnoidal wave solution (right). The reference waveform is constructed using the same effective coefficients and wavelength as in the simulation, enabling a like-for-like comparison. The sustained periodicity and stable waveform confirm that under weakly nonlinear, long-wavelength conditions, cnoidal waves closely match their KdV counterparts.}
\label{fig:cnoidal}
\end{figure}

\begin{figure}[ht!]
\centering
\includegraphics[width=0.45\textwidth]{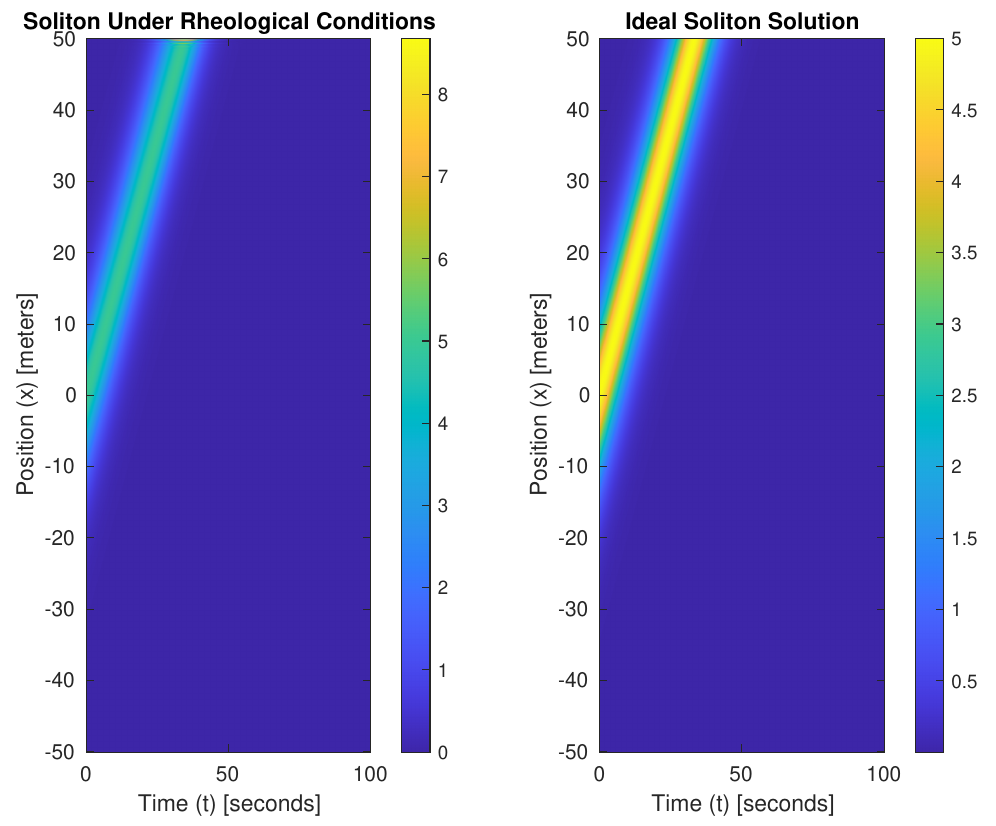}
\caption{Soliton wave propagation under rheological conditions (left) compared to the ideal KdV soliton (right). After a brief spin-up, the pulse translates with nearly constant speed and width, and the early transient radiation remains weak. The soliton maintains its localized shape and speed, demonstrating robust agreement with the KdV prediction.}
\label{fig:soliton}
\end{figure}

Figures~\ref{fig:cnoidal} and \ref{fig:soliton} compare the numerically simulated waveforms (left panels) with the corresponding ideal KdV solutions (right panels) for cnoidal and soliton waves, respectively. The numerical solutions were obtained with the conservative finite-volume solver described above and assessed with phase-speed, amplitude-drift, and shape-misfit diagnostics; across all runs shown, these diagnostics remain small over many wave periods, consistent with weakly nonlinear, long-wave propagation.

For the cnoidal cases, the simulated waveform retains its period and crest-to-trough structure over the full integration; slight amplitude relaxation and a gentle phase downshift are attributable to basal dissipation and the regularized Bingham closure, not to dispersive error. The KdV reference provides an accurate envelope for the resolved wave over many periods.

For the solitary-pulse cases, the misfit quickly plateaus after the initial adjustment and remains nearly constant thereafter, indicating asymptotic propagation with minimal deformation.

Together, these comparisons support the applicability of a KdV balance for fluidized debris-flow surges within the admissible corridor identified earlier; outside that corridor, such as steep reaches where roll-wave dynamics dominate, the KdV model is not invoked.

\subsection{Energy and Momentum Evolution of Cnoidal Waves}

\begin{figure}[h!]
\centering
\includegraphics[width=0.95\columnwidth]{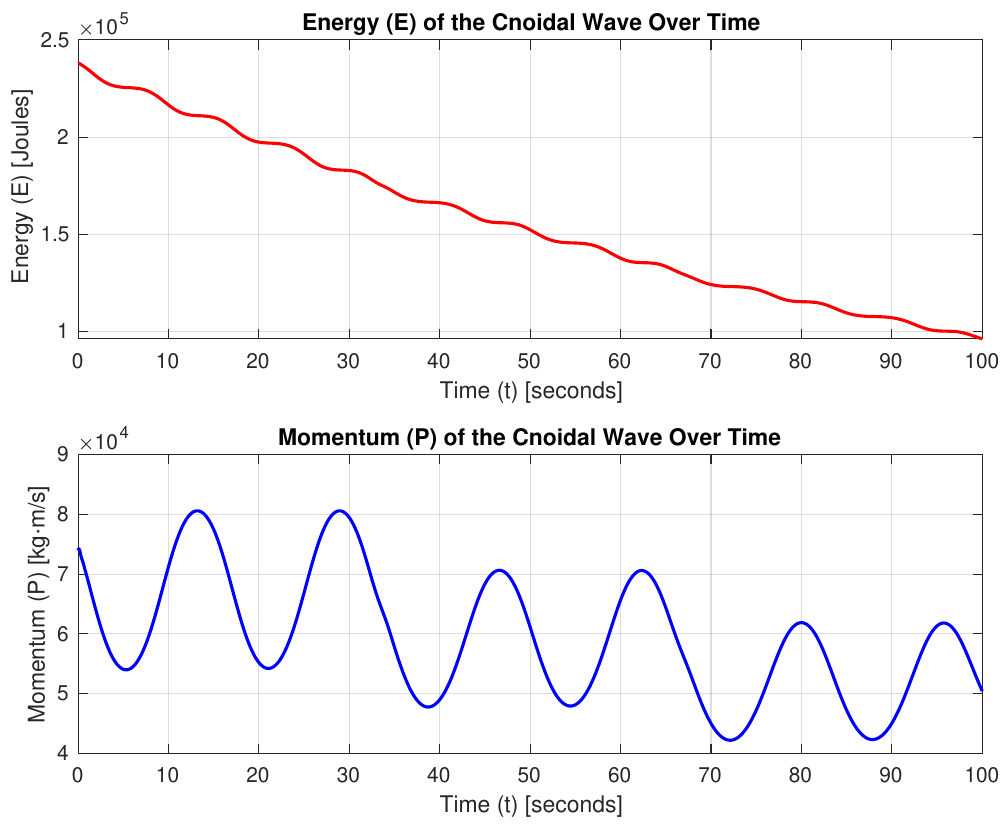}
\caption{Temporal evolution of energy ($E$) and momentum ($P$) for a cnoidal wave under rheological conditions. Both quantities decrease over time, with oscillations indicating internal energy redistribution rather than purely monotonic decay, consistent with dissipative surge-front behavior. Curves are domain-integrated diagnostics computed consistently with the numerical fluxes under periodic boundaries; the gentle secular decline reflects basal work and rheological dissipation, while the small oscillations at the fundamental and first harmonic indicate periodic steepening and relaxation rather than numerical leakage.}
\label{fig:EP_vs_time}
\end{figure}

Throughout, when quoting or plotting mechanical energy $E$ and horizontal momentum $P$ for a single, steadily translating pulse, we enforce the mechanical consistency $E=\tfrac{1}{2}Pc$, with $c$ the tracked crest speed. To quantify persistence under the full rheology, we track depth-integrated mechanical energy $E(t)$ and horizontal momentum $P(t)$ computed consistently with the finite-volume fluxes and source terms. For cnoidal runs, the time series exhibit a slow, monotonic drift due to basal work, superposed on small oscillations at the fundamental and first harmonic that reflect periodic steepening and relaxation. This behavior is expected for weakly nonlinear waves propagating in a weakly dissipative medium and aligns with the small phase-speed and shape-misfit errors reported by the diagnostics. Surge-front waves with steep, quickly moving leading edges of high friction and locally high coarse-material concentration~\cite{Iverson1997,takahashi1978mechanical} can possess periodic structures like cnoidal waves under certain circumstances. Figure~\ref{fig:EP_vs_time} shows that $E(t)$ and $P(t)$ decrease gradually while exhibiting periodic oscillations, reflecting dissipative dynamics rather than a loss of coherence.

In our simulations, the initial energy is approximately $2.4 \times 10^5 \text{J}$ and decreases to about $1.0 \times 10^5 \text{J}$ at $t = 100 \text{s}$. This corresponds to a reduction of roughly $58\%$, computed from $(2.4-1.0)/2.4 \approx 0.58$. The decline reflects basal work and rheological dissipation (viscous losses and yielding) rather than numerical leakage, which aligns with frictional losses expected for fluidized debris flows. The oscillatory trend is indicative of internal energy redistribution---a feature of nonlinear systems---and shows that coherence degrades gradually without external forcing, even though the carrier waveform remains recognizable over many periods.

Similarly, the momentum $P(t)$ decreases from roughly $7.5 \times 10^4 \text{kg}\cdot\text{m/s}$ to about $5.0 \times 10^4 \text{kg}\cdot\text{m/s}$ over the same period, with a peak-to-peak variation of nearly $2.5 \times 10^4 \text{kg}\cdot\text{m/s}$. This is an $\approx 33\%$ net reduction, with the superposed oscillations indicating front-tail exchanges of momentum typical of weakly dispersive, weakly nonlinear trains.

The mass/impulse proxy is
\begin{equation}
\mathcal{P} = \int_{-\infty}^{\infty} \eta(x,T_{\mathrm{exit}}) dx,
\label{eq:momentum_def}
\end{equation}
where $T_{\mathrm{exit}}$ is the diagnostic time: for solitary‐pulse runs it is chosen just before the crest reaches the outflow boundary (at least one pulse width upstream), and for periodic cnoids it is an integer number of wave periods after spin-up. In practice the integral is evaluated over the numerical domain with a composite-trapezoid rule. The subscripts “rheo” and “ideal” denote the value computed from the full rheological simulation and from the matched ideal KdV waveform (constructed with the same $(h_0,u_0,\alpha,\beta)$ and amplitude), respectively.

The normalized score is
\begin{equation}
M_p = \max \left(0,  1 - \frac{\lvert \mathcal{P}_{\mathrm{rheo}}-\mathcal{P}_{\mathrm{ideal}}\rvert}
{\lvert \mathcal{P}_{\mathrm{ideal}}\rvert + \epsilon_{\mathrm{tol}}}\right).
\label{eq:Mp}
\end{equation}
Here $\epsilon_{\mathrm{tol}}=10^{-6}$ is a small numerical tolerance (distinct from the asymptotic small parameter $\epsilon$) that prevents division by zero and sets a floor. With this choice, the invariant-based scores $E_p$ and $M_p$ both lie in $[0,1]$. These proxies are invariants only for inviscid KdV; in the full rheological model they serve solely as fidelity diagnostics, and we compute them for both solitons and cnoids.

\subsection{Energy and Momentum Evolution of Soliton Waves\label{sec:soliton_energy}}

\begin{figure}[h!]
\centering
\includegraphics[width=\columnwidth]{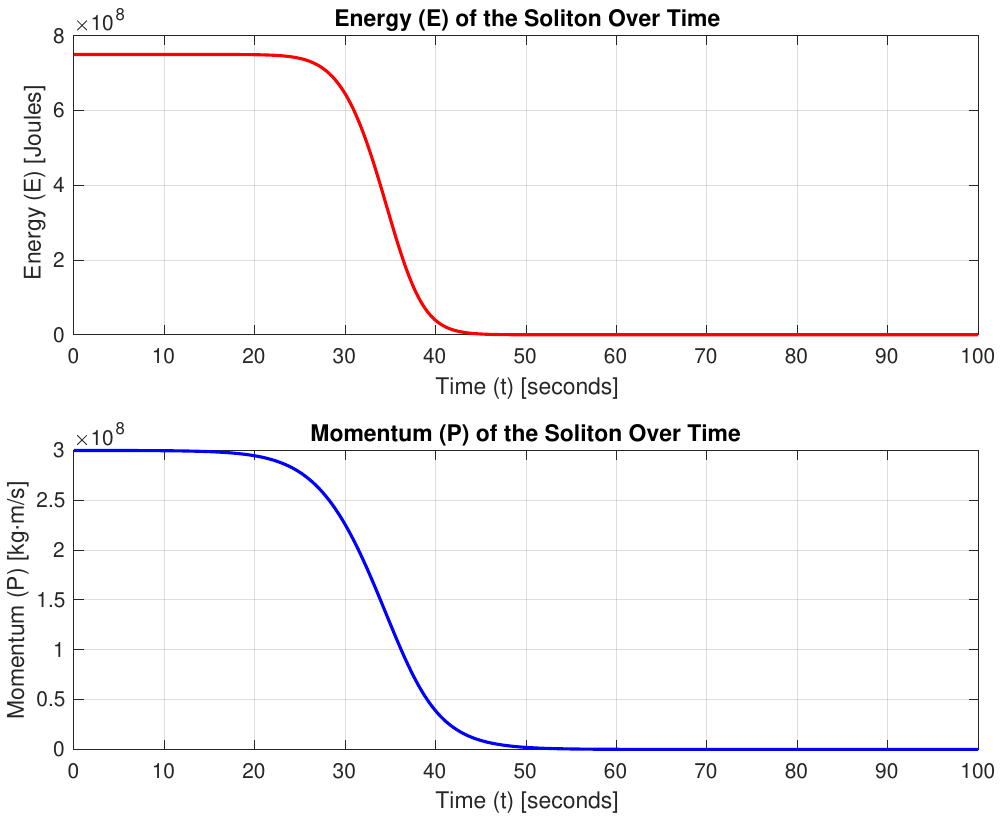}
\caption{Temporal evolution of energy $E$ and momentum $P$ for the soliton under the full depth-averaged rheological model. While the pulse remains within the computational domain, both quantities vary only weakly; once it exits through the downstream free-exit boundary they rapidly approach zero, illustrating its role as a momentum-carrying solitary pulse. Small oscillations superposed on the plateaus reflect weak radiation and reversible exchange with the dispersive tail rather than numerical leakage.}
\label{fig:soliton_EP_vs_time}
\end{figure}

We next examine how a solitary wave’s energy $E(t)$ and momentum $P(t)$ evolve under the full depth-averaged rheological model, interpreting these results in the context of surge front waves---the steep, fast-moving leading edges of debris flows \cite{Iverson1997, Hungr2005}. In the simulations, the soliton is initialized with a $\operatorname{sech}^2$-type profile and allowed to propagate freely through the computational domain, mimicking the concentrated energy and momentum transfer typically observed in surge fronts. Figure~\ref{fig:soliton_EP_vs_time} illustrates the temporal evolution of $E$ and $P$. Here $E$ and $P$ denote the channel-integrated kinetic energy and streamwise momentum of the excess thickness $\eta=h-h_0$ (units J and $\text{kg}\cdot\text{m/s}$); for a steadily translating pulse with crest speed $V$ one has $E \approx \tfrac12 P V$, so these quantities are directly comparable to the distal work $|W|$ (J).

Consistent with the weakly dissipative setting, the time series show slow, nearly monotonic decay of both energy and momentum while the waveform preserves its localized envelope and translational speed to within numerical tolerance. Small, superposed oscillations arise from weak radiation shed during spin-up and reversible exchanges between the core of the pulse and its dispersive tail, rather than from numerical leakage.

Relative to the cnoidal case, the soliton exhibits markedly smaller attenuation over the same window, which we attribute to its compact structure and reduced interior shear. In the conservative KdV limit, $E$ and $P$ would be invariants; here, departures from constancy are expected because finite yield stress $\tau_y$ and depth-averaged viscous dissipation $\mu_{\mathrm{eff}}$ inject physical damping absent from the inviscid theory. This behavior supports the use of solitary pulses as efficient carriers of momentum through low-slope reaches, provided the weak-amplitude and long-wave assumptions hold.

To connect with the macroscopic budget, we interpret the instantaneous loss rate $-\mathrm{d}E/\mathrm{d}t$ of a propagating pulse as the mechanical power expended against basal resistance along the reach. For an upper-bound, energy-based admissibility test we compare the energy available from a pulse (or a train of pulses) with the distal work magnitude $|W|$ defined in Eq.~(\ref{eq:work}). Because Eq.~(\ref{eq:work}) uses $|\tau_d-\tau_r|$, all benchmarks are made against $|W|$, so cases with $\tau_r>\tau_d$ on very gentle slopes are handled consistently.

The parameter map in Fig.~\ref{fig:energy} reports $|W|$ only; it does not assume any particular number of pulses. Here $N$ denotes the number of pulses in the train. We deem a train energetically sufficient if
\begin{equation}
N E_{\mathrm{soliton}} \gtrsim |W|,  \quad E_{\mathrm{soliton}}=\tfrac{1}{2} P_{\mathrm{soliton}} V,
\end{equation}
for a representative pulse with crest speed $V$. An equivalent power view integrates the loss rate $-\mathrm{d}E/\mathrm{d}t$ over a traverse time $T \approx L/V$ to give the same benchmark.

Across the ranges used in Fig.~\ref{fig:energy}, the simulated loss rates and representative pulse energies imply $N$ within the observed range, indicating that soliton trains can plausibly offset the distal basal work over distances comparable to the field cases. This mechanism is regime-selective: it is most effective for small $\epsilon=a/h_0$, moderate base Froude number, and slopes within or below the dispersive corridor, and it need not occur in every debris-flow setting.

During the early phase the soliton maintains nearly constant total energy and momentum. Using the representative pulse speed in these runs ($c \approx 10~\text{m s}^{-1}$), we find
$E \approx 2.0\times 10^{6}\,\text{J}$ and $P \approx 4.0\times 10^{5}\,\text{kg}\cdot\text{m/s}$, which satisfy $E=\tfrac{1}{2} P c$.
These values indicate that the pulse preserves the balance between nonlinear steepening and dispersive spreading characteristic of surge fronts. Subsequently both $E$ and $P$ decrease as the pulse leaves the domain through the downstream free-exit boundary; this drop reflects boundary flux rather than intrinsic dissipation, implying that a soliton-like surge front can propagate effectively until it encounters external boundaries or additional loss mechanisms.

This sustained plateau in energy and momentum underscores the robustness of the solitary pulse even in the presence of depth-averaged viscosity $\mu_{\mathrm{eff}}$ and modest nonlinear steepening. Here, nonlinear steepening is quantified by $\alpha A$ via the KdV speed–amplitude relation $V = u_0 + c_0 + (\alpha A)/3$ (see Sec.~\ref{sec:DebrisFlowAnalysis}); distinct from this, $\gamma_{\mathrm{nl}}$ is the dimensionful (s$^{-1}$) nonlinearity parameter used only in the numerical fidelity survey (see the subsection~\ref{sec:FidelityPropagatingSoliton}, “Fidelity of Propagating Soliton” and Fig.~\ref{fig:fidelity_plot}). Under suitable conditions—weak nonlinearity, long wavelengths, and moderate slopes—the pulse transports energy and momentum over appreciable distances, supporting its use as an idealized model for momentum-carrying surge fronts in debris flows.

The close agreement between the theoretical benchmarks and the numerically tracked $E(t)$ and $P(t)$ supports the use of KdV model as a first-order description of surge-front dynamics in fluidized debris flows while emphasizing that applicability depends on the stated asymptotic and slope constraints.

\subsection{Fidelity of Propagating Soliton}
\label{sec:FidelityPropagatingSoliton}

To quantify agreement between the numerically simulated solitary pulse and an ideal KdV pulse, we use a composite fidelity that combines waveform similarity with two conservation proxies based on KdV invariants. The proxies use the squared-amplitude integral and the area under the free surface; they are dimensionless and are not the mechanical energy and momentum used for runout energetics.

The waveform score $W_s$ is a normalized $L_2$ discrepancy at the exit time $T_{\mathrm{exit}}$:
\begin{equation}
W_s = \max \left(0, 
1 - 
\frac{\bigl\|\eta_{\mathrm{rheo}}(\cdot,T_{\mathrm{exit}})-\eta_{\mathrm{ideal}}(\cdot,T_{\mathrm{exit}}) \bigr\|_{2}}
{\bigl\| \eta_{\mathrm{ideal}}(\cdot,T_{\mathrm{exit}}) \bigr\|_{2}+\epsilon}
\right),
\label{eq:Ws}
\end{equation}
which ensures $0\le W_s\le 1$ and reduces sensitivity to absolute amplitude. In practice, norms are evaluated with a composite-trapezoid rule on the numerical grid.

The quadratic invariant is
\begin{equation}
\mathcal{E} = \int_{-\infty}^{\infty} \eta^{2}(x,T_{\mathrm{exit}}) dx,
\label{eq:energy_def}
\end{equation}
and its score is
\begin{equation}
E_p = \max \Bigl(0, 
1 - \frac{\bigl| \mathcal{E}_{\mathrm{rheo}}-\mathcal{E}_{\mathrm{ideal}}\bigr|}
{\mathcal{E}_{\mathrm{ideal}}+ \epsilon_{\mathrm{tol}}}\Bigr).
\label{eq:Ep}
\end{equation}

The mass/impulse proxy $M_p$ is defined in Eq.~\eqref{eq:Mp} using
$\mathcal{P}=\int \eta(x,T_{\mathrm{exit}}) dx$ from Eq.~\eqref{eq:momentum_def}.

The overall fidelity is the unweighted mean
\begin{equation}
F = \frac{W_s + E_p + M_p}{3}, \quad F\in[0,1],
\label{eq:F}
\end{equation}
which summarizes the proximity of the rheological solution to the ideal KdV pulse at $T_{\mathrm{exit}}$.

\begin{figure}[h!]
\centering
\includegraphics[width=0.9\columnwidth]{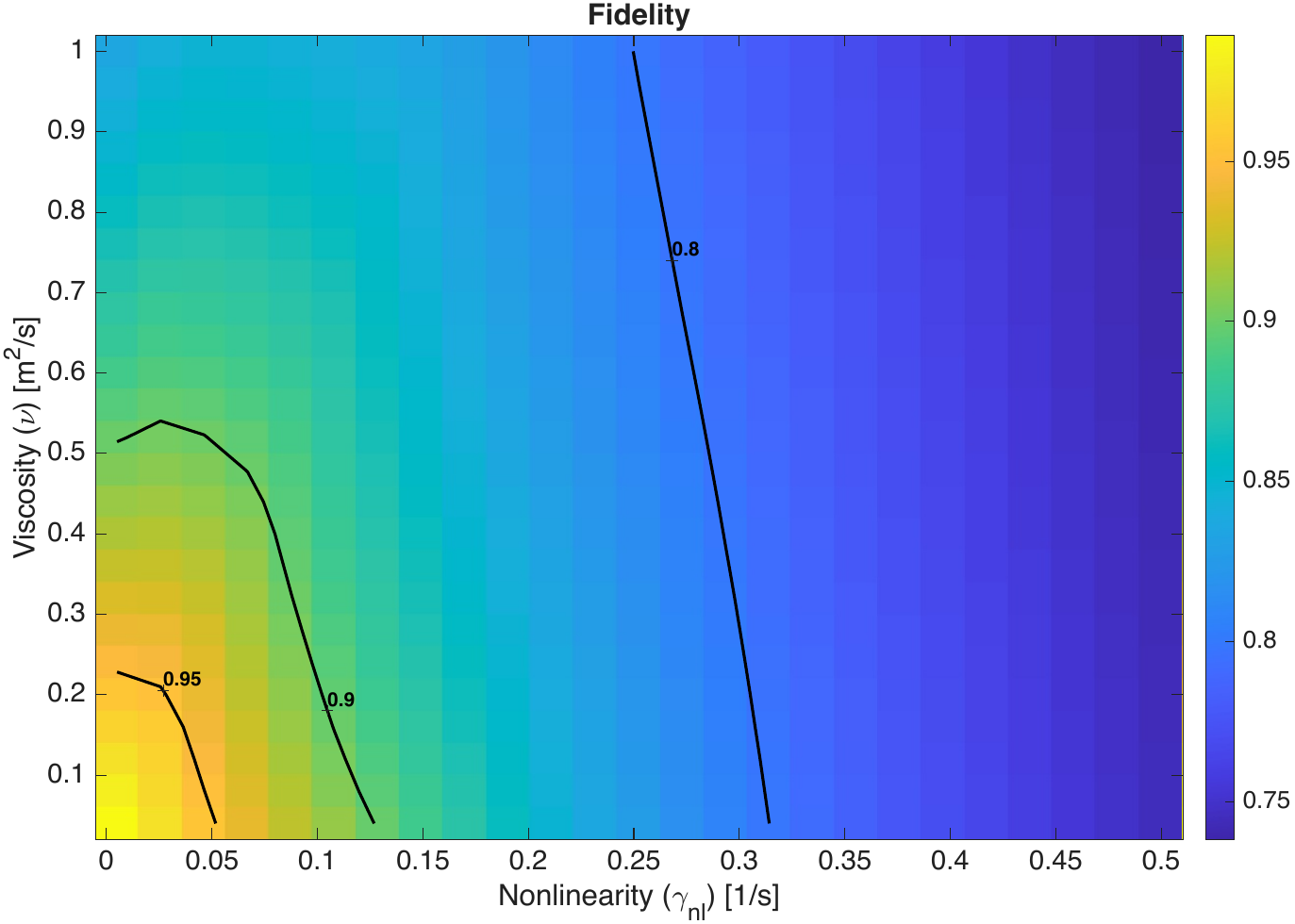}
\caption{\textbf{Fidelity} $F\in[0,1]$ as a function of the nonlinearity parameter $\gamma_{\mathrm{nl}}$ (s$^{-1}$) and the bulk kinematic viscosity $\nu$ (m$^{2}$ s$^{-1}$). Contours mark $F=0.95$, $0.90$, and $0.80$. High fidelity occurs for small $\nu$ and $\gamma_{\mathrm{nl}}$; fidelity decreases as either parameter increases. Here $\nu$ is the survey’s bulk kinematic viscosity, distinct from the depth-averaged effective viscosity $\mu_{\mathrm{eff}}$ (Pa$\cdot$s/m) and the Coulomb friction coefficient $\mu_s$ (dimensionless).}
\label{fig:fidelity_plot}
\end{figure}

We surveyed two solver controls: a nonlinearity parameter $\gamma_{\mathrm{nl}}$ (s$^{-1}$) and a bulk kinematic viscosity $\nu$ ($\text{m}^{2}\text{ s}^{-1}$). Here $\gamma_{\mathrm{nl}}$ multiplies the quadratic nonlinearity in the numerical model and sets an effective nonlinearity timescale; it is not the curvature–stiffness factor $\gamma$ used in Sec.~\ref{sec:dispersion} to set dispersion (there $\beta_{\mathrm{eff}}=(c_0 h_0^2/6)\gamma$ with $\gamma$ dimensionless). In the survey, we held $\gamma>0$ (and thus $\beta_{\mathrm{eff}}$) fixed and varied $0.005\le \gamma_{\mathrm{nl}}\le 0.5~\text{s}^{-1}$ and $0.04\le \nu \le 1.0~\text{m}^{2}\text{ s}^{-1}$\footnote{Here $\nu$ is a numerical bulk kinematic viscosity. It is distinct from the depth-averaged effective viscosity $\mu_{\mathrm{eff}}$ (Pa$\cdot$s/m) used in the rheological model and from the Coulomb friction coefficient $\mu_s$ (dimensionless).}.
For each pair, we computed $F$ and contoured $F=0.95, 0.90, 0.80$ (Figure~\ref{fig:fidelity_plot}).

High $F$ occurs for small $\nu$ and modest $\gamma_{\mathrm{nl}}$. Fidelity decreases as either parameter increases. Within the plotted range, nonlinearity degrades $F$ more rapidly than viscosity. These patterns are consistent with weakly nonlinear, weakly dissipative propagation of solitary pulses in the long-wave regime.

High fidelity ($F>0.95$) is obtained only when both $\gamma_{\mathrm{nl}}$ and $\nu$ are small, indicating weak nonlinearity and low dissipation. As $\gamma_{\mathrm{nl}}$ increases, $F$ decreases sharply. Across most of the grid, the $F=0.8$ contour lies near $\gamma_{\mathrm{nl}}\approx 0.3~\text{s}^{-1}$, which suggests that nonlinearity degrades soliton persistence more than viscosity within the explored ranges. These patterns are consistent with prior analyses \cite{InfeldRowlands2000, Ancey2007}.

The composite metric $F$ is therefore a compact diagnostic for soliton persistence under parameter variation. In field settings with heterogeneity, three-dimensional geometry, and time-dependent rheology, the admissible region in the $(\gamma_{\mathrm{nl}},\nu)$ plane is expected to be narrower. Additional processes such as entrainment and pore-pressure evolution will likely shift the contours toward lower $F$; quantifying these shifts requires site-specific constraints.

\section{Discussion}

\subsection{Soliton Fidelity and Normalized Nonlinearity}

We show that depth-averaged debris-flow equations augmented with non-Newtonian rheology can admit traveling waves that closely resemble KdV solutions under weak-nonlinearity, long-wavelength conditions. In water-saturated, fluidized debris flows these coherent waves provide a simplified surrogate for surge fronts that convey energy and momentum over distance. By contrast, for large and mainly dry landslides, other mechanisms (lubrication, air cushioning, dynamic fragmentation) are likely to dominate, so a KdV description becomes less appropriate.

A key outcome is the fidelity metric $F$ that quantifies how closely numerically simulated solitons track their ideal KdV counterparts. High fidelity occurs only when dissipation and the nonlinearity parameter remain modest, consistent with soliton-like surge fronts in fluid-rich flows. In contrast, cnoidal waves exhibit greater dissipation and reduced persistence in our full rheological simulations.

We also use a normalized nonlinearity measure to organize regimes by base-flow Froude number and to identify parameter windows in which KdV-type assumptions plausibly hold. Taken together, the fidelity metric and the normalized nonlinearity provide complementary diagnostics for when solitary-pulse transport is dynamically admissible.

\subsection{Model Limitations and Future Outlook}

The shallow, weakly nonlinear, long-wave reduction is most credible in tail reaches or other low-to-moderate slopes where fluidization persists and curvature-induced normal stresses contribute dispersion. Many documented events descend steeper terrain before entering runout zones; mean fan gradients in our compilation frequently exceed ten percent, with local gentler segments possible along portions of the path. Accordingly, KdV-type pulses should be interpreted as a conditional pathway—not a universal outcome—activated where fluid support, moderate slopes, and weak nonlinearity coexist.

Additional physics—pore-pressure transients, entrainment, turbulence, and strong grain-size segregation—can add dissipation beyond the present closure. Extending the model to include these processes may improve runout and impact predictions while retaining solitary pulses as computationally efficient surrogates for surge fronts.

Future work should address higher-order asymptotics, two- and three-dimensional geometries, and time-dependent rheology. Systematic use of the normalized nonlinearity together with the base-flow Froude number can guide experiment and field campaigns toward conditions where coherent pulses are most likely, and where the KdV approximation can be tested rigorously.

\subsection{Next-Order Corrections to the KdV Equation}

As shown in Figure~\ref{fig:fidelity_plot}, the leading-order KdV approximation has a finite domain of validity. Consider the one-dimensional depth-averaged system
\begin{align}
\frac{\partial h}{\partial t} + \frac{\partial}{\partial x}(h u) &= 0,\\
\frac{\partial (h u)}{\partial t} + \frac{\partial}{\partial x}\Bigl(h u^2 + \tfrac{1}{2}g h^2\Bigr) 
&= -\tau_b -\frac{\partial \tau_{xx}}{\partial x}, \nonumber
\end{align}
where $h=h(x,t)$ is the flow depth, $u=u(x,t)$ the depth-averaged velocity, $g$ gravitational acceleration, $\tau_b$ basal resistance, and $\tau_{xx}$ an internal stress that supplies dispersion. Suppose
\begin{align}
h(x,t) &= h_0 + \epsilon h_1(x,t) + \epsilon^2 h_2(x,t) + \cdots,\nonumber\\
u(x,t) &= u_0 + \epsilon u_1(x,t) + \epsilon^2 u_2(x,t) + \cdots,
\end{align}
with $\epsilon\ll1$ measuring weak nonlinearity ($a/h_0$) and scale separation ($h_0/L$). At $\mathcal{O}(\epsilon)$ one recovers the KdV balance
\begin{equation}
\eta_t + c \eta_x + \alpha \eta \eta_x + \beta \eta_{xxx} = 0,
\end{equation}
where $\eta\propto h_1$, and $c$ is the linear wave speed ($c=u_0+c_0$ in the laboratory frame or $c=c_0$ in a frame moving with $u_0$, with $c_0$ as defined in §~\ref{sec:notation}); $\alpha$ and $\beta$ depend on the chosen closure for $\tau_b$ and $\tau_{xx}$.

For more energetic waves or shorter length scales, the $\mathcal{O}(\epsilon^2)$ terms cannot be neglected. Retaining second-order contributions from $h u^2$, $(g/2)h^2$, $\partial_x\tau_{xx}$, and the rheology yields a solvability condition that augments KdV with cubic nonlinearity and higher-order dispersion:
\begin{equation}
\eta_t + c \eta_x + \alpha \eta \eta_x + \beta \eta_{xxx}
+ \gamma_2 \eta^{2} \eta_x + \beta_2 \eta_{xxxxx} = 0.
\end{equation}
Here the coefficients depend on the second-order structure of the momentum flux and normal stresses; the cubic term modifies steepening at larger amplitudes, whereas the fifth derivative controls short-wavelength dispersion and front structure. These corrections help bridge the gap between high-fidelity solitary pulses and regimes where the leading-order approximation begins to fail.

\section{Conclusion}

Under weak nonlinearity and long wavelengths, a depth-integrated debris-flow model with non-Newtonian rheology reduces to a KdV-type description that admits both periodic (cnoidal) and solitary (soliton) waves. In water-rich, fluidized debris flows, these coherent structures represent dispersive pulses in the fines-rich tail; solitons act as localized carriers of energy and momentum that can modulate motion on gentle reaches, not the debris-flow front itself.

Natural surges also exhibit a high-friction head and a more fluid, viscous tail, and processes such as entrainment and pore-pressure transients can add dissipation beyond the present closure, so our results should be viewed as a conditional description rather than a full predictive model. The fidelity metric quantifies when simulated solitary pulses remain close to their ideal KdV counterparts, highlighting that stability is sustained only when dissipation and effective nonlinearity are modest. A velocity-based nonlinearity proxy, $\tilde{\alpha}_{\mathrm{norm}}$, plotted against the base-flow Froude number $\mathrm{Fr}_0$, provides a practical regime diagnostic that separates subcritical from supercritical conditions and indicates where KdV-type assumptions are most plausible; where crest celerity is available we compare with $\alpha_{\mathrm{norm}}$.

Future extensions should incorporate higher-dimensional geometry, time-dependent rheology, and entrainment, and should be tested against targeted laboratory and field measurements to confirm the occurrence and persistence of soliton-like dispersive pulses in fines-rich tails under natural conditions. By coupling classical nonlinear-wave theory with depth-averaged rheological modeling, this study clarifies when a balance of nonlinearity and dispersion can sustain coherent pulses and indicates how such pulses may contribute to momentum transport across gentle reaches over long distances.

\appendix

\section{Nomenclature (List of symbols)}
\label{sec:nomenclature}

\begin{table}[h!]
\centering
\caption{List of symbols used in this work.\label{tab:nomenclature}}
\begin{tabular}{@{}llp{5.2cm}@{}}
\toprule
Symbol & Units & Meaning \\
\midrule
$h(x,t)$ & m & Flow depth; $h_0$ base depth \\
$u(x,t)$ & m/s & Depth-averaged velocity; $u_0$ base velocity \\
$\zeta=h-h_0$ & m & Dimensional free-surface perturbation \\
$\eta=\zeta/h_0$ & -- & Dimensionless free-surface perturbation \\
$u'=u-u_0$ & m/s & Velocity perturbation; $\hat u=u'/c_0$ \\
$c_0=\sqrt{g h_0}$ & m/s & Linear shallow-water speed \\
$\alpha_0=3c_0/(2h_0)$ & s$^{-1}$ & Canonical quadratic nonlinearity \\
$\beta_0=c_0 h_0^2/6$ & m$^3$/s & Canonical dispersion coefficient \\
$\alpha,\beta$ & s$^{-1}$, m$^3$/s & Effective KdV coefficients \\
$\gamma$ & -- & Curvature–stiffness factor in $\tau_{xx}$ closure \\
$\tau_{xx}$ & Pa & Internal normal (dispersive) stress \\
$\tau_b$ & Pa & Basal shear stress (Coulomb or viscous–plastic) \\
$\mu_s$ & -- & Coulomb friction coefficient (tail: $\mu_{\mathrm{tail}}$) \\
$\eta_{\mathrm{3D}}$ & Pa·s & Intrinsic dynamic viscosity (3D) \\
$\mu_{\mathrm{eff}}=\eta_{\mathrm{3D}}/h$ & Pa·s/m & Depth-averaged effective viscosity \\
$S=\tan\theta$ & -- & Bed slope \\
$\mathrm{Fr}=u/\sqrt{g h}$ & -- & Froude number; base value $\mathrm{Fr}_0$ \\
$A$ & m & Wave amplitude ($h_{1s}-h_0$) \\
$V$ & m/s & Crest celerity (lab frame) \\
$L_{\mathrm{w}}$ & m & Wave horizontal scale (wavelength or soliton width) \\
$L$ & m & Geometric reach length (for work budget) \\
$W$ & J & Distal work magnitude $= \rho g h_{\mathrm{tail}} W_b L^2 |S-\mu_{\mathrm{tail}}|$ \\
$w$ & m & Along-crest (flow) width \\
$W_b$ & m & Channel width used in distal-work budget \\
$A_{\mathrm{cs}}$ & m$^2$ & Cross-sectional area (parameter estimate) \\
\bottomrule
\end{tabular}
\end{table}

\section{Energy and Momentum of a Soliton in a 3D Debris Flow}\label{sec:AppendixA}

This appendix derives depth-integrated expressions for the kinetic energy and streamwise momentum carried by a solitary elevation (soliton) in a debris flow, treating the wave as a shallow free-surface pulse of excess thickness $\eta(x,t)$ that spans an along-crest planform width $w$ (for channel-filling pulses, $w\approx W_b$, the channel width used in the main text). The results quantify the transport capacity of a coherent pulse under the weakly nonlinear, long-wave assumptions used in the main text, and neglect potential-energy changes and internal strain-energy corrections.

\subsection{Derivation of Energy and Momentum}

We assume the soliton is described by the canonical profile
\begin{equation}
\eta(x,t) = A \sech^2 \bigl[ \kappa (x - c t)\bigr], 
\end{equation}
where $A$ (m) is the amplitude, $\kappa$ (m$^{-1}$) is the inverse characteristic length scale $L_s=\kappa^{-1}$ (half-width at half-maximum $=0.881 L_s$), and $c$ ($\text{m s}^{-1}$) is the constant propagation speed (here $c$ denotes the crest speed $V$). Let $\rho$ ($\text{kg m}^{-3}$) denote bulk density and $w$ (m) the along-crest flow width.

Under a shallow, depth-averaged description, the excess mass per unit streamwise length associated with the pulse is $\rho w \eta(x,t)$. We define the kinetic energy $E$ and streamwise momentum $P$ of this soliton as
\begin{equation}
\begin{aligned}
E &= \frac{1}{2} \int_{-\infty}^{\infty} \rho w \eta(x,t) c^{2} dx,\\
P &= \int_{-\infty}^{\infty} \rho w \eta(x,t) c dx .
\end{aligned}
\end{equation}
Here $E$ is the mechanical kinetic energy of the excess mass translating at speed $c$ (neglecting base KE and potential/strain terms), distinct from the KdV invariant $\int \eta^{2} dx \sim \frac{4 A^{2}}{3\kappa}$ for elevation.
Assuming $\rho$ and $w$ are constant along the soliton, these simplify to
\begin{equation}
E = \frac{1}{2}\rho w c^2 \int_{-\infty}^{\infty} \eta(x,t) dx,
\quad P = \rho w c \int_{-\infty}^{\infty} \eta(x,t) dx.
\end{equation}
Substituting the soliton profile into $E$ and $P$ gives
\begin{align}
E =& \frac{1}{2}\rho w c^2 A
\int_{-\infty}^{\infty} \sech^2\bigl[\kappa (x - c t)\bigr] dx,\nonumber \\
P =& \rho w c A
\int_{-\infty}^{\infty} \sech^2\bigl[\kappa (x - c t)\bigr] dx.
\end{align} 
Using $u = \kappa(x - c t)$ with $dx = du/\kappa$ and $\int_{-\infty}^{\infty} \sech^2(u) du = 2$, we find
\begin{equation}
E = \frac{\rho w A c^2}{\kappa}, \quad P = \frac{2 \rho w A c}{\kappa}.
\end{equation}
Equivalently, the soliton’s excess mass and the implied energy and momentum are
\begin{equation}
\begin{aligned}
M_{\eta} &= \rho w \int \eta(x,t) dx = \frac{2 \rho w A}{\kappa},\\
P &= M_{\eta} c,\quad E = \frac{1}{2} M_{\eta} c^{2},
\end{aligned}
\end{equation}
which confirms dimensional and mechanical consistency.

\subsection{Estimating Soliton Energy and Momentum in Debris Flows}

As an order-of-magnitude illustration, take the following representative values for debris-flow material and wave parameters:
\begin{equation}
\begin{aligned}
\rho &= 2000~\text{kg m}^{-3},&
w &= 10~\text{m},&
c &= 10~\text{m s}^{-1},\\
A &= 5~\text{m},&
\kappa &= \tfrac{1}{5}~\text{m}^{-1}.
\end{aligned}
\end{equation}
(so the characteristic length is $L_s=\kappa^{-1}$; the half-width at half-maximum is $\operatorname{arcosh}(\sqrt{2}) L_s \approx 0.881 L_s$)
\begin{equation}
L_s = \kappa^{-1} = 5~\text{m}.
\end{equation}
From the soliton energy formula we obtain
\begin{equation}
E = \frac{\rho w A c^{2}}{\kappa}
= \frac{2000\times 10 \times 5 \times 10^{2}}{1/5}
= 5\times 10^{7}~\text{J}.
\end{equation}
Similarly, from the momentum expression
\begin{equation}
P = \frac{2\rho w A c}{\kappa}
= \frac{2 \times 2000 \times 10 \times 5 \times 10}{1/5}
= 1\times 10^{7}~\text{kg}\cdot\text{m/s}.
\end{equation}
These satisfy $E=\tfrac{1}{2}Pc$ as required. In applications, $c$ should be taken as the observed surge (or pulse) speed, and the formulas applied to the elevation component $\eta\ge 0$; departures from the $\sech^{2}$ shape, vertical structure, or lateral variability can be accommodated via the measured $\int \eta dx$.

\section{Estimating the Resistive Forces Opposing Soliton-Driven Propulsion of the Debris Flow Wavefront}

The forward motion of a debris flow is primarily driven by the component of gravity acting downslope, $M g \sin\theta$. Nonetheless, several resistive forces can significantly diminish the net downslope momentum, including basal friction, cohesive yield stress, internal shear, and pore-pressure effects \cite{iverson1997debris,hungr2005estimating,takahashi1978mechanical,rotaru2007analysis,zhao2022flexible,pudasaini2024mechanically}. Among these, basal friction often dominates, especially in surge fronts---the steep, fast-moving leading edges distinguished by high friction and a coarse grain composition~\cite{Iverson1997, Hungr2005, takahashi1978mechanical}.

\subsection{Coulomb basal-friction variant}
\label{app:coulomb}

Field observations show that basal resistance is often dominated by Coulomb friction rather than a pressure-independent yield stress~\cite{Iverson1997,Hungr2005}. We model the basal shear as
\begin{equation}
\tau_b = \mu_s \rho g h \cos \theta \operatorname{sgn}(u),
\label{eq:coulomb_basal}
\end{equation}
with constant friction coefficient $\mu_s$. For a uniform base state with $u_0>0$, the steady downslope balance is $\rho g \sin \theta = \mu_s \rho g \cos \theta$, i.e.\ $\tan \theta= \mu_s$. Under the weakly nonlinear, long-wave ordering $\epsilon \sim(h_0/L_{\mathrm{w}})^2 \ll1$ used throughout, the multiple-scale expansion about this balanced base has the following consequence: because $\tau_b$ in \eqref{eq:coulomb_basal} is rate-independent and $\operatorname{sgn}(u)$ does not change for small perturbations around $u_0>0$, the Coulomb term cancels with the slope at leading order and does not appear in the $\mathcal{O}(\epsilon)$ reduced dynamics.

Eliminating the velocity perturbation with the linear slaving relation, the laboratory-frame elevation obeys, to $\mathcal{O}(\epsilon)$,
\begin{equation}
\eta_t + (u_0+c_0) \eta_x + \alpha_{\mathrm{eff}} \eta \eta_x + \beta_{\mathrm{eff}} \eta_{xxx} = 0,
\label{eq:kdv_coulomb_leading}
\end{equation}
where $c_0$ is as defined in §~\ref{sec:notation}, with the same $\alpha_{\mathrm{eff}}$ and $\beta_{\mathrm{eff}}$ as in the viscous-plastic baseline. In particular, Coulomb friction does not renormalize the quadratic nonlinearity at this order (no factor of the form $\alpha\mapsto \alpha \Xi(\mu_s)$ in the leading KdV balance).

Basal Coulomb work is nevertheless accounted for in the full-order solver via its standard mechanical power,
\begin{equation}
\dot{E}_b = -\int \tau_b u dx = - \rho g \mu_s\cos\theta \int h u dx,
\label{eq:basal_work_coulomb}
\end{equation}
which contributes to attenuation in the fully resolved energetics. Under the asymptotic assumptions above (balanced base, no sign change), this basal term produces no $\mathcal{O}(\epsilon)$ correction to \eqref{eq:kdv_coulomb_leading}; any Coulomb influence appears only beyond leading order (e.g., through departures from exact base balance or local sign changes), and is treated numerically rather than by introducing ad hoc damping in the analytical KdV model.

The leading-order cancellation requires a steady, sign-definite base current $u_0>0$ and $\tan\theta=\mu_s$; local reversals, spatial variability of $\mu_s$, or base-state mismatch break the cancellation and yield a small dissipative correction beyond the leading-order KdV balance.

\subsection{Parameter Estimation}

We provide an order-of-magnitude estimate of the resistive forces that a solitary surge front must overcome under representative field conditions.
Adopt the following nominal values (chosen to reflect coarse, surge-front material): yield stress $\tau_y$, cross-sectional area $A_{\mathrm{cs}}$, basal friction coefficient $\mu_s$ (Coulomb), density $\rho$, gravity $g$, slope angle $\theta$, control-volume length $L$, and channel width $w$:
\begin{align}
\tau_y &= 500~\text{Pa},            & A_{\mathrm{cs}} &= 10~\text{m}^2,      & \mu_s &= 0.30, \nonumber\\
\rho   &= 2000~\text{kg m}^{-3},    & g &= 9.81~\text{m s}^{-2}, & \theta &= 5^\circ, \nonumber\\
L      &= 10~\text{m},              & w &= 10~\text{m}.        & \nonumber
\end{align}
The mass of a segment of length $L$ is
\begin{equation}
M = \rho A_{\mathrm{cs}} L = 2000 \times 10 \times 10 = 2.0\times 10^{5}\ \text{kg}.
\end{equation}
The cohesive (yield) contribution acts over the basal (planform) area $A_{\text{basal}} = w L$ rather than the cross-sectional area. Taking $w=10~\text{m}$ gives $A_{\text{basal}} = 10 \times 10 = 100~\text{m}^2$, so
\begin{equation}
F_{\text{yield}} = \tau_y A_{\text{basal}} = 500 \times 100 = 5.0\times 10^{4}\ \text{N}.
\end{equation}
The Coulomb basal-friction contribution is
\begin{align}
F_{\text{friction}} =&\ \mu_s M g \cos\theta \nonumber \\
=&\ 0.30 \times \bigl(2.0 \times 10^{5}\bigr) \times 9.81 \times \cos 5^\circ, \nonumber \\
\approx&\ 5.86\times 10^{5}\ \text{N}.
\end{align}
Hence the total resistive force is
\begin{align}
F_{\text{total}} =&\ F_{\text{yield}} + F_{\text{friction}}  \nonumber \\
\approx&\ (5.0\times 10^{4}) + (5.86\times 10^{5}) \nonumber \\
\approx&\ 6.36\times 10^{5}\ \text{N}.
\end{align}

For context, the downslope driving component on the same control volume is
\begin{align}
F_{\text{drive}} = M g \sin\theta
&= \bigl(2.0\times 10^{5}\bigr) \times 9.81 \times \sin 5^\circ, \nonumber \\
&\approx 1.71\times 10^{5}\ \text{N}.
\end{align}
At these representative values, the resistive demand exceeds the downslope driving by a factor of about $F_{\text{total}}/F_{\text{drive}}\approx 6.36\times 10^{5}/(1.71\times 10^{5}) \approx 3.7$, implying that a solitary front would decelerate unless basal resistance is reduced (e.g., by elevated pore pressure, lubrication, or diminished coarse packing).

Comparing $F_{\text{total}}$ with the energy and momentum carried by the soliton (see Sec.~\ref{sec:soliton_energy}) helps assess whether the wave can maintain its surge-front character over meaningful distances. These numbers are illustrative and can be updated straightforwardly with site-specific parameters to refine the assessment.

\subsection{Soliton Energy and Momentum Analysis}

We estimate the soliton momentum as
\begin{equation}
P_{\mathrm{soliton}} \approx 4 \times 10^{5}\ \text{kg}\cdot\text{m/s},
\end{equation}
which imparts a velocity increment to a debris segment of mass $M=2\times 10^{5}\ \text{kg}$:
\begin{equation}
\delta v = \frac{P_{\mathrm{soliton}}}{M}
= \frac{4 \times 10^{5}}{2 \times 10^{5}}
\approx 2\ \text{m s}^{-1}.
\end{equation}
Here $\delta v$ denotes a velocity change (not a grid spacing).  This $\mathcal{O}(1\text{--}2)\ \text{m s}^{-1}$ boost is consistent with surge fronts exceeding the mean base-flow speed.

For comparison with resistive demands, recall from the previous subsection that the total basal resistance for the control volume is
\begin{equation}
F_{\mathrm{total}} \approx 6.36 \times 10^{5}\ \text{N}.
\end{equation}
Using a one-meter displacement as a convenient benchmark,
\begin{equation}
E_{\mathrm{diss}}(d{=}1\ \text{m}) = F_{\mathrm{total}} d
\approx 6.36 \times 10^{5}\ \text{J}.
\end{equation}
Taking a representative soliton speed $c \approx 10\,\text{m s}^{-1}$, the corresponding energy (enforcing $E=\tfrac{1}{2}P_{\mathrm{soliton}}c$) is
\begin{equation}
E_{\mathrm{soliton}} = \tfrac{1}{2} P_{\mathrm{soliton}} c
= \tfrac{1}{2} (4\times 10^{5})\times 10
= 2.0\times 10^{6}\ \text{J}.
\end{equation}
Therefore,
\begin{equation}
\frac{E_{\mathrm{soliton}}}{E_{\mathrm{diss}}}
\approx \frac{2.0 \times 10^{6}}{6.36 \times 10^{5}}
\approx 3.14.
\end{equation}
Thus, per meter of advance, the solitary pulse carries a factor of $\approx 3.1$ more energy than the local basal work at these nominal values; sustained long runout still requires favorable conditions (e.g., reduced effective friction, transient pore-pressure support) and/or gravitational input along the reach.

In short, the illustrative magnitudes $E_{\mathrm{soliton}}\sim 2\times 10^{6}\ \text{J}$ and $P_{\mathrm{soliton}}\sim 4\times 10^{5}\ \text{kg}\cdot\text{m/s}$ (with $c\approx 10\ \text{m s}^{-1}$) satisfy $E=\tfrac{1}{2}Pc$ and indicate that solitary pulses can plausibly sustain surge-front motion against basal friction and cohesion locally, while acknowledging that additional dissipation (internal shear, turbulence, entrainment) will erode this budget over distance.

{\bf Acknowledgments:} We thank William I. Newman and David Jewitt for helpful discussions.
This work is supported by the National Aeronautics and Space Administration under Grant/Contract/Agreement No. 80NSSC24K0671 through the Earth Surface and Interior Program.

\bibliographystyle{apsrev4-2}
\bibliography{refs}

\end{document}